\documentclass[twocolumn,pra,aps,english,floatfix,a4paper]{revtex4-1} 
\usepackage[english]{babel}
\usepackage{amsmath,amssymb}
\usepackage{graphicx,tikz,epsfig,graphics,subfigure}
\usepackage{amssymb}
\usepackage{amsmath}
\usepackage{commath}
\usepackage{graphicx,bm}
\usepackage{verbatim}
\usepackage{enumitem}
\usepackage{listings}
\usepackage[capitalise]{cleveref}
\usepackage{multirow}

\Crefname{appsec}{App.}{Apps.}
\crefname{appsec}{App.}{Apps.}

\usepackage{booktabs}

\AtBeginDocument{
	\heavyrulewidth=.08em
	\lightrulewidth=.05em
	\cmidrulewidth=.03em
	\belowrulesep=.65ex
	\belowbottomsep=0pt
	\aboverulesep=.4ex
	\abovetopsep=0pt
	\cmidrulesep=\doublerulesep
	\cmidrulekern=.5em
	\defaultaddspace=.5em
}

\renewcommand{\vec}{\mathbf}
\newcommand{\kvec}{{\vec{k}}}
\newcommand{\Rvec}{{\vec{R}}}
\newcommand{\tvec}{{\vec{t}}}
\newcommand{\ket}[1]{{\left|#1\right\rangle}}
\newcommand{\angstrom}{\textup{\AA}}

\definecolor{green}{RGB}{00,136,51}
\definecolor{blue}{RGB}{0,51,153}
\definecolor{orange}{RGB}{238,102,0}
\lstdefinestyle{pythonCode}{
  language=Python,
  commentstyle=\color{green}\fontseries{b}\fontshape{sl}\selectfont,
  keywordstyle=\color{blue}\fontseries{b}\selectfont,
  stringstyle=\color{orange},
  frame=both,
  tabsize=4,
}
\begin{document}

\title{Automated construction of symmetrized Wannier-like tight-binding models from \textit{ab initio} calculations}
\author{Dominik Gresch$^1$}
\author{{QuanSheng} Wu$^2$}
\thanks{The work was carried out at Theoretical Physics, ETH Zurich, 8093 Zurich, Switzerland}
\author{Georg W. Winkler$^{3}$}
\thanks{The work was carried out at Theoretical Physics, ETH Zurich, 8093 Zurich, Switzerland}
\author{Rico H\"auselmann$^{4}$}
\author{Matthias Troyer$^{1,5}$}
\author{Alexey A. Soluyanov$^{1,6,7}$}
\affiliation{${^1}$Theoretical Physics, ETH Zurich, CH-8093 Zurich, Switzerland}
\affiliation{${^2}$Institute of Physics, \'{E}cole Polytechnique F\'{e}d\'{e}rale de Lausanne (EPFL), CH-1015 Lausanne, Switzerand}
\affiliation{${^3}$Station Q, Microsoft Research, Santa Barbara, California 93106-6105, USA}
\affiliation{${^4}$Theory and Simulation of Materials (THEOS) and National Centre for Computational Design and Discovery of Novel Materials (MARVEL), \'Ecole Polytechnique F\'ed\'erale de Lausanne, 1015 Lausanne, Switzerland}
\affiliation{${^5}$Microsoft Research, Redmond, WA, United States}
\affiliation{${^6}$St. Petersburg State University, St. Petersburg, 199034, Russia}
\affiliation{${^7}$Physik-Institut, Universit\"at Z\"urich, Winterthurerstrasse 190, CH-8057 Zurich, Switzerland}

\begin{abstract}
Wannier tight-binding models are effective models constructed from first-principles calculations. 
As such, they bridge a gap between the accuracy of first-principles calculations and the 
computational simplicity of effective models. In this work, we extend the existing methodology of creating Wannier tight-binding models 
from first-principles calculations by introducing the symmetrization post-processing step, which enables the production of Wannier-like models that 
respect the symmetries of the considered crystal. Furthermore, we implement automatic workflows, which allow for producing a large number of tight-binding models for large classes of chemically and structurally similar compounds, or materials subject to external influence such as strain.
As a particular illustration, these workflows are applied to strained III-V semiconductor materials. These results can be used for further study of topological phase transitions in III-V quantum wells.
\end{abstract}

\maketitle
\section{Introduction}

A significant part of materials science is devoted to the problem of finding the electronic structure of a given material. As a result, numerous computational techniques have been developed to study this problem. These techniques can roughly be classified into two kinds: \emph{First-principles} methods solve the problem using the fundamental physical principles and properties of atoms comprising the material. For weakly-interacting systems, density functional theory (DFT)~\cite{hohenberg_kohn_dft} is the dominant (mean field) technique for solving the electronic structure problem from first principles. 

In contrast, \emph{empirical} methods aim to capture the relevant physical properties using a simplified model. Such models are usually matched to known properties of the material, which can be obtained from either experiments or first-principles calculations. An example of such an empirical method is given by the tight-binding approximation, which describes a material as a set of localized orbitals and predefined electron hopping terms between them. While the first-principles methods typically have superior accuracy, empirical methods are often used due to their lower computational cost. In particular, calculations of complex device geometries are often inaccessible to a direct first-principles study. As such, the construction of reliable empirical models is of significant importance. And the technique of creating Wannier tight-binding models~\cite{marzariMaximallyLocalizedGeneralized1997,souzaMaximallyLocalizedWannier2001} from first-principles calculations is arguably one of the most popular tools in nowadays computational materials science. The use of Wannier tight-binding models allows one to combine the simplicity of empirical methods with the correct wave function properties obtained from first-principles.

In recent years, \emph{high-throughput} techniques made a profound impact in various fields of materials science~\cite{franceschettiInverseBandstructureProblem1999,johannessonCombinedElectronicStructure2002,curtaroloPredictingCrystalStructures2003,curtaroloHighthroughputHighwayComputational2013}. While the domain eludes a strict definition, a common feature of such techniques is that computational tools are applied to a wide range of candidate materials, or variations of a given material, in search of some beneficial property. Existing codes and techniques are combined and applied on a scale that was not previously possible. A range of automated frameworks~\cite{aiida, jainFireWorksDynamicWorkflow2015} support this by facilitating the combination of separate calculations into logical workflows. The challenge in designing such a high-throughput workflow is to make it resilient to varying input parameters. Since the number of calculations performed is too large to be human-controlled, many decisions -- for example which calculation to perform based on the output of a previous calculation -- need to be encoded into the automated workflow.

In this paper, we introduce steps for addressing two standardly known problems of using Wannier90~\cite{wannier90, wannier90_updated} in combination with any \textit{ab initio} software to construct tight-binding models: the absence of symmetries present in the original compound in the obtained tight-binding model, and the neccessity to search for optimal inner and outer energy windows for projection of the first-principles energy bands. 
We do not, however, treat the issue of selecting the initial projections used by Wannier90. 
As such, we create automated workflows which are applicable to large classes of materials with similar orbital character of the bands of interest. However, these workflows are not yet applicable to high-throughput scenarios in the sense that they can trivially be applied to arbitrary compounds. 
Nevertheless, the presented workflows are written in a way that they could be combined with efforts to address the problem of selecting initial projections~\cite{mustafaAutomatedConstructionMaximally2015}.

In Sec.~\ref{sec:wannier_tb}, we review the general process of calculating the Wannier tight-binding models by means of Wannier90 and explain the proposed and implemented symmetrization and automatic energy window choice procedures. Sec.~\ref{sec:implementation} describes how these procedures are used for the development of an automated workflow using the {AiiDA}~\cite{aiida} framework. While this workflow automates the tight-binding calculation itself, there are still some tunable parameters which might be eliminated by a more sophisticated system. By using a modular design approach, we provide an extensible framework for implementing such improvements. In the final section, we illustrate the application of this workflow to calculate tight-binding models for strained III-V semiconductor materials. These are useful in the pursuit of Majorana devices~\cite{kitaevUnpairedMajoranaFermions2001,lutchynMajoranaFermionsTopological2010,oregHelicalLiquidsMajorana2010}, enabling the study of transport properties for different topological devices with III-V quantum wells, where strains play an important role in the topological transition.

\section{Construction of Wannier-like tight-binding models}
\label{sec:wannier_tb}

In this section, we describe the process of generating \textit{symmetrized} Wannier-like tight-binding (SWTB) models. First, we give a short description of the method for creating Wannier tight-binding models (WTB) as introduced in the works of Refs.~\cite{marzariMaximallyLocalizedGeneralized1997,souzaMaximallyLocalizedWannier2001} and implemented in the Wannier90~\cite{wannier90,wannier90_updated} software package. Next, we describe a method for symmetrizing these WTBs in a post-processing step. Finally, we describe a scheme to enhance the band-structure accuracy by \textit{optimizing the energy windows} used by Wannier90.

\subsection{Wannier tight-binding construction}

Tight-binding models represent a common way to describe crystalline systems in a computationally cheap way. The material is described as a system of localized orbitals with positions $\tvec_i$ in the unit cell, and hopping terms $H^{ij}[\Rvec]$ between the $j$-th orbital in the unit cell at location $\Rvec$ and the $i$-th orbital in the home unit cell $\Rvec={\bf 0}$. From these parameters, the matrix Hamiltonian can be written as~\footnote{In this work, we use the tight-binding convention I of Ref.~\cite{pythtb_formalism}.}
\begin{equation}\label{eqn:tight_binding}
\mathcal{H}^{ij}(\kvec) = \sum_\Rvec H^{ij}[\Rvec]e^{i \kvec.(\Rvec + \tvec_j - \tvec_i)}.
\end{equation}
For the case of spinful systems, we choose the indices $i, j$ to include the spin index for simplicity.

The Wannier tight-binding (WTB) method utilizes localized Wannier functions as basis orbitals to capture the compound's physics. These basis Wannier functions are obtained from first-principles simulations. This procedure is based on the work of Refs.~\cite{marzariMaximallyLocalizedGeneralized1997,souzaMaximallyLocalizedWannier2001} and implemented in the Wannier90~\cite{wannier90,wannier90_updated} code. After obtaining the necessary Wannier90 input files from a first-principles calculation, two steps are performed to construct these Wannier functions:

In a first step, the Bloch wave-functions $\ket{\psi_{n, \kvec}}$ calculated by the first-principles code are \emph{disentangled} to obtain $M$ wave-functions, where $M$ is the target number of basis Wannier functions in WTB. For selecting the Bloch wave-functions which are involved in this procedure, one needs to choose an \emph{outer} energy window. Optionally, an \emph{inner} energy window can be chosen. States inside this inner window will be preserved by the disentanglement. An optimization routine is performed to select the $M$ states such that the ``change of character'' $\Omega_\text{I}$ (defined in Ref.~\cite{souzaMaximallyLocalizedWannier2001}) is minimized. As an initial guess for this optimization procedure, $M$ localized trial orbitals $\ket{g_m}$ are used. Because the disentanglement procedure needs to discard some states, it usually changes both the symmetry and the energy bands of the model in comparison with first-principles results. Consequently, choosing good values for both the energy windows and the trial orbitals has a strong effect on the quality of the resulting model.

As a second (optional) step, another optimization is performed to find a unitary transformation such that the resulting Wannier functions are maximally localized~\cite{marzariMaximallyLocalizedGeneralized1997}. Again, the trial orbitals $\ket{g_m}$ are used to create an initial guess for this optimization. Typically, these orbitals are chosen to be those chemical atomic orbitals that contribute most to the bands of interest. A method for constructing Wannier orbitals without the need for such a guess is described in Ref.~\cite{mustafaAutomatedConstructionMaximally2015}.

\subsection{Symmetrization}

An important feature of tight-binding models, especially for studying topological effects, is that they preserve certain crystal symmetries. For a given symmetry group $G$, the symmetry constraint on the Hamiltonian matrix is given by~\cite{dresselhausGroupTheoryApplication2007}
\begin{equation}\label{eqn:symmetry_constraint}
\forall g \in G: \mathcal{H}(\vec{k}) = D^\kvec(g) \mathcal{H}(g^{-1}\vec{k}) D^\kvec(g^{-1}),
\end{equation}
where $D^\kvec(g)$ is the $\kvec$-dependent representation of the symmetry $g$ from the group $G$. We define the $\kvec$ - \emph{independent} part $D(g)$ of the representation as 
\begin{equation}
D^\kvec(g) = e^{i \boldsymbol{\alpha}_g.\kvec} D(g),
\end{equation}
where $\boldsymbol{\alpha}_g$ is the translation vector of the symmetry.

For a Hamiltonian which does not fulfill these symmetry constraints, we define the \emph{symmetrized} Hamiltonian as the group average
\begin{equation}\label{eqn:symmetrized_hamiltonian}
\tilde{\mathcal{H}}(\vec{k}) = \frac{1}{|G|} \sum_{g \in G} D^\kvec(g) \mathcal{H}(g^{-1}\vec{k}) D^\kvec(g^{-1}).
\end{equation}
This procedure projects the Hamiltonian onto the symmetric subspace, meaning that the modified Hamiltonian respects \cref{eqn:symmetry_constraint}, as shown in \cref{appendix:symmetrized_tb}. Furthermore, if the original Hamiltonian is already symmetric, the original and symmetrized Hamiltonians are identical. Since this construction does not explicitly construct the corresponding Wannier functions, we term these models symmetrized \emph{Wannier-like} tight-binding models (SWTB).

It is important to note that the eigenstates and eigenvalues of the symmetrized Hamiltonian may differ significantly from those of the non-symmetrized Hamiltonian. In fact, for an anti-symmetric initial Hamiltonian, meaning that
\begin{equation}
D^\kvec(g)\mathcal{H}(g^{-1}\kvec)D^\kvec(g^{-1}) = - \mathcal{H}(\kvec)
\end{equation} 
for some symmetry $g$, the symmetrized result vanishes completely. However, given a Hamiltonian which \emph{almost} respects the symmetry, this technique can effectively eliminate small symmetry-breaking terms.

In the context of tight-binding models, this symmetrization technique can only straightforwardly be applied when the underlying basis set is symmetric. If the tight-binding basis contains an orbital $\ket{\alpha}$ centered around the position $\vec{r}$, it must also contain $g\ket{\alpha}$ centered around $g \vec{r}$ for all symmetries $g \in G$. For example, if the model for a material which has $C_4^x$ symmetry contains a $p_x$ orbital at the origin, it must also contain a $p_y$ orbital at the origin.

For Wannier tight-binding models, this means that the technique can generally only be applied when the step of maximally localizing the Wannier functions is omitted, and pre-defined atomic orbitals are used.
When this condition is met however, the method can be applied for both unitary and anti-unitary symmetries, as well as non-symmorphic symmetry groups.

To apply the group average to tight-binding models, it is convenient to rewrite Eq.~\ref{eqn:symmetrized_hamiltonian} directly in terms of the hopping matrices $H[\vec{R}]$ (see App.~\ref{app:symmetrized_H_in_real_space} for derivation):
\begin{equation}
\tilde{H}^{ij}[\Rvec] = \frac{1}{|G|} \sum_{\substack{g \in G \\{l,m}}} D_{il}(g) H^{lm}[S_g^{-1}(\Rvec - \vec{T}_{ij}^{ml})] D_{mj}(g^{-1}),
\end{equation}
where $S_g$ is the real-space rotation matrix of the symmetry $g$, $\vec{T}_{ij}^{ml} = S_g(\tvec_m - \tvec_l) - \tvec_j - \tvec_i$, and the indices $m, l$ only go over values for which $\vec{T}_{ij}^{ml}$ is a lattice vector. Note that we use the $\kvec$ - independent part $D(g)$ of the representation here.

\begin{figure*}\centering
\includegraphics[width=\textwidth]{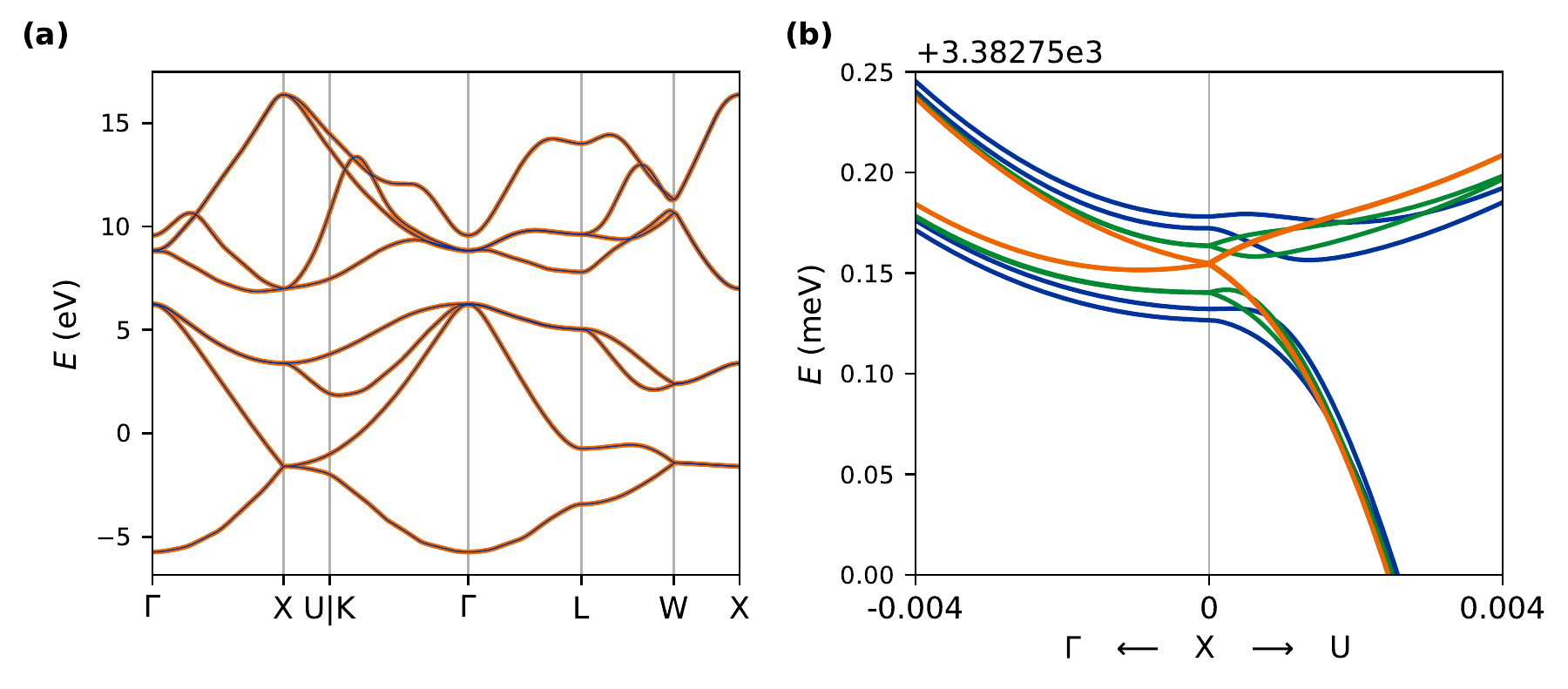}
\caption{(Color online) Comparison of the initial (blue) and symmetrized (orange) bandstructure for a tight-binding model of silicon with atom-centered $sp^3$ orbitals. (a) In the $\mathrm{eV}$ scale, there are no visible differences between the two models. (b) A zoom in around the $X$ point on the $\mathrm{meV}$ scale reveals a slight lifting of the band degeneracies in the initial model. This incorrectness is resolved in the symmetrized model. For comparison, a symmetrized bandstructure taking into account only symmorphic symmetries (green) is also shown.}
\label{fig:symmetrization}
\end{figure*}

Fig.~\ref{fig:symmetrization} shows the results of this symmetrization procedure on a tight-biding model for bulk silicon in the diamond cubic crystal structure, with atom-centered $sp^3$ orbitals. The initial model already approximately fulfills the symmetry condition, which is reflected in the fact that the bandstructure does not change in the electronvolt scale. However, at the sub-millielectronvolt scale the band degeneracies are lifted in the original model, but restored after the symmetrization procedure. Since the symmetry group of the diamond cubic structure $\mathrm{Fd}\bar{3}\mathrm{m}$ (no. $227$) is non-symmorphic, this example demonstrates that the symmetrization technique is capable also of enforcing such symmetries. In panel b of \cref{fig:symmetrization}, we compare the symmetrization using the full symmetry group to a partial symmetrization enforcing only the symmorphic subgroup. Adding non-symmorphic symmetries enforces the four-fold degeneracy at the $\mathrm{X}$ point and two-fold degeneracy on the $\mathrm{X} - \mathrm{U}$ line, whereas symmorphic symmetries only enforce a two-fold degeneracy on the $\Gamma - \mathrm{X}$ line.

To determine the matrix representations $D(g)$, we use the fact that Wannier90 allows one to manually choose the trial orbitals $\ket{g_m}$. As a result, the basis after the disentanglement procedure corresponds to the chosen orbitals, up to some numerical error. Since the behavior of the basis orbitals under symmetries is known, $D(g)$ can be determined in this way. For the treatment of spin, we use the rotation matrices as given in ref.~\cite{haberThreeDimensionalProperImproper2011}. The action of time-reversal on the spin basis $\{\ket{\uparrow}, \ket{\downarrow} \}$ is given by $\sigma_y \hat{K}$, where $\hat{K}$ represents complex conjugation.
An automated method for generating the representation matrices for given atomic orbitals is available in the \texttt{symmetry-representation} package. 
Importantly, we used Wannier90 \textit{without} performing the maximal localization step. It is the case in the illustrated application of Sec.~\ref{sec:strain_models}, where this allows us to preserve the orbital basis. Alternatively, one could use the basis transformation matrices $U^{(\kvec)}$ provided by Wannier90~\cite{wannier90} to transform $D(g)$ into the maximally-localized basis. While this approach produces computationally cheaper localized models, the drawback is that the basis is different for each produced tight-binding model. As a result, comparing models is more difficult. Also, linear interpolation between models, as described in \cref{sec:strain_interpolation}, would require a change of basis.

Another approach to obtaining symmetric tight-binding models is to use the site-symmetry mode implemented in Wannier90~\cite{sakumaSymmetryadaptedWannierFunctions2013}. However, this method is limited to symmetries which leave a given real-space coordinate invariant (site symmetries), and does not include time-reversal. The method presented here has no such limitation, but is instead limited to models which have a symmetric set of basis functions as described above. The site-symmetry mode also relies on obtaining the symmetry information from the first-principles code, which is currently implemented only for Quantum Espresso~\cite{espresso,espresso_updated}. The workflow described in Sec.~\ref{sec:aiida_workflows} could be adapted to allow using this approach with only minimal changes.

\subsection{Optimization for bandstructure fit}

As described above, an important parameter in running Wannier90 is the choice of the so-called energy windows~\cite{wannier90}. There are two such windows: The \emph{outer} window determines which states are taken into account for the disentanglement procedure. At every $\kvec$-point, it must contain at least $M$ bands, where $M$ is the desired number of bands in the tight-binding model. The \emph{inner} (or frozen) window on the other hand determines which states should not be modified during disentanglement. It can contain at most $M$ bands at any given $\kvec$.

Since the quality of the resulting tight-binding model depends sensitively on the choice of energy windows, a strategy for reliably choosing good windows is required.
A straightforward way of achieving this is by iteratively optimizing the window values. Having constructed and symmetrized a tight-binding model, its quality can be determined by comparing its bandstructure to a reference computed directly from first-principles \footnote{Because the first-principles calculation usually contains more than $M$ bands, we need to choose which bands should be represented by the tight-binding model.}. As a measure of their mismatch, we choose the average difference between the energy eigenvalues
\begin{equation}
\Delta = \frac{1}{M}\frac{1}{N_\kvec} \sum_{i=1}^{M}\sum_\kvec \left| \varepsilon_{i, \kvec}^\text{DFT} - \varepsilon_{i, \kvec}^\text{TB} \right|.
\label{eqn:average_band_difference}
\end{equation}
Some values of the energy windows cannot produce a tight-binding model, for example if the outer window contains less than $M$ bands. As a result, finding appropriate energy windows is a constrained, four-dimensional optimization problem. The Nelder-Mead (downhill simplex) algorithm~\cite{nelderSimplexMethodFunction1965} can be used to solve this problem~\footnote{The constraint is implemented by assigning an infinite value of $\Delta$ to invalid energy windows.}.

\begin{figure*}\centering
\includegraphics[width=\textwidth]{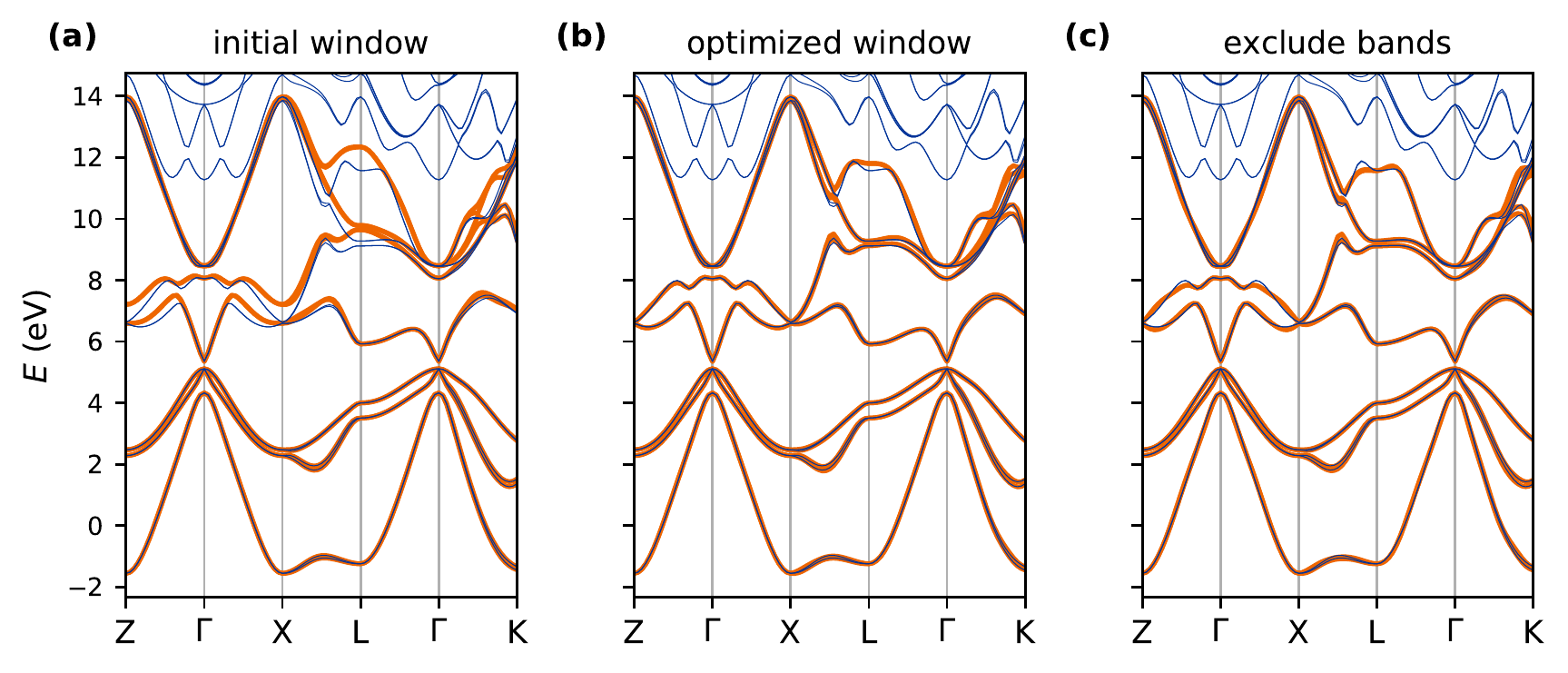}
\caption{(Color online) Comparison between the reference first-principles bandstructure (blue) and bandstructures calculated from tight-binding models (orange) for InSb. The tight-binding model in (a) was calculated with the initial energy window, whereas (b) shows the model using the optimized energy window as detailed in \cref{tab:energy_window_opt}. The model in (c) was calculated without the disentanglement procedure, using the \texttt{exclude\_bands} parameter.}
\label{fig:optimize_energy_window}
\end{figure*}

Fig.~\ref{fig:optimize_energy_window} shows the result of such an optimization procedure for unstrained InSb, as described in \cref{sec:strain_models}. A clear improvement is visible between the tight-binding model obtained with the initial windows chosen by hand (panel a), and the optimized window values (panel b). In particular, the conduction bands at the $X$ and $Z$ points are represented more accurately in the optimized model. Since the given bands for InSb are not entangled, it is also possible to skip the disentanglement step completely by using the \texttt{exclude\_bands} parameter of Wannier90 to ignore all other energy bands. The resulting bandstructure is shown in \cref{fig:optimize_energy_window}(c). Nevertheless, we find that the bandstructure using optimized disentanglement is slightly better ($\Delta=0.0327$) than the one without disentanglement ($\Delta=0.0375$), especially for the four lowest conduction bands on the $\mathrm{Z}$ - $\Gamma$ - $\mathrm{X}$ line.
Hence, it can be useful to apply the disentanglement procedure and energy window optimization even in cases where the bands are not inherently entangled, especially when the time required to run the tight-binding calculation is short compared to the initial first-principles calculation.

\section{Implementation in A\lowercase{ii}DA workflows}\label{sec:aiida_workflows}
\label{sec:implementation}

The AiiDA \cite{aiida} platform is a Python framework for performing high-throughput calculations, focused on the field of materials physics. It enables reproducible research by keeping track of inputs, outputs and settings for each calculation. On top of this provenance layer, it provides a toolset for automatically chaining calculations into user-defined workflows.

In this section, we describe the implementation of the Wannier tight-binding extraction scheme as an AiiDA workflow. This automation enables the application to the study of strain effects (described in \cref{sec:strain_models}). Special care has been taken to design the workflow in a modular way, which enables re-using parts of the workflow for purposes other than tight-binding extraction. We first discuss these design principles, before showing how they are applied in the tight-binding workflows.

The code for the AiiDA workflows is available in the open-source \texttt{aiida-tbextraction} package, and provided as supplementary material.

\subsection{Modular workflow design}

The basic principle of modular workflow design is to split up a single monolithic workflow into minimal sub-workflows or calculations that perform exactly one task. For example, the tight-binding model created by Wannier90 is post-processed by parsing it to an HDF5 format, followed by optionally changing the order of the basis and symmetrizing the model. While this could easily be implemented in a single script, splitting these three steps up into separate calculations allows separately re-using each of the steps.

More complex workflows are created by combining multiple sub-workflows into a logical unit at a higher abstraction level. Inputs to the sub-workflow are either forwarded directly from the input to the parent workflow or created within the parent workflow. Similarly, outputs from the sub-workflow can either be forwarded to be an output of the parent workflow or consumed directly to guide the further execution of the parent workflow.

Since a complex workflow can consist of multiple layers of wrapped sub-workflows, this modular approach is maintainable only if the overhead of forwarding input and output is minimal. Following the single responsibility principle, a parent workflow should not have to change if an input or output parameter of a sub-workflow changes, unless it directly interacts with this parameter. To achieve this, a syntax is needed to specify that a parent workflow will \emph{inherit} inputs or outputs of a sub-workflow, without explicitly listing each parameter.
In AiiDA, such a feature is available in the newly-introduced \emph{expose} functionality, as described in \cref{app:expose}.

The modular architecture improves not only the re-usability, but also the flexibility of workflows. Often, a given part of a workflow could be performed in different ways. For example, many different codes can perform the first-principles calculations in the tight-binding extraction workflows. Additionally, one might want to add steps such as relaxation or cut-off energy convergence.

To allow for this, the parent workflow can allow for dynamically selecting a workflow for performing a given task by passing it as an input~\footnote{For storing the workflow in the AiiDA database, it needs to be converted into an AiiDA data type. We chose to convert it into a string containing the fully qualified class name, from which we import  the workflow when needed.}. An abstract workflow class defines the interface that a workflow must fulfill so that it can be used to perform the task. If needed, the parent workflow can allow for dynamic inputs, which are just forwarded to the specific workflow implementing the interface. In this way, the parent workflow can act as a template that defines an abstract series of steps, without knowledge of the detailed input flags available on each step.

\subsection{Tight-binding extraction workflow}\label{sec:tb_workflow}

\begin{figure}\centering
\includegraphics[width=0.8844\columnwidth]{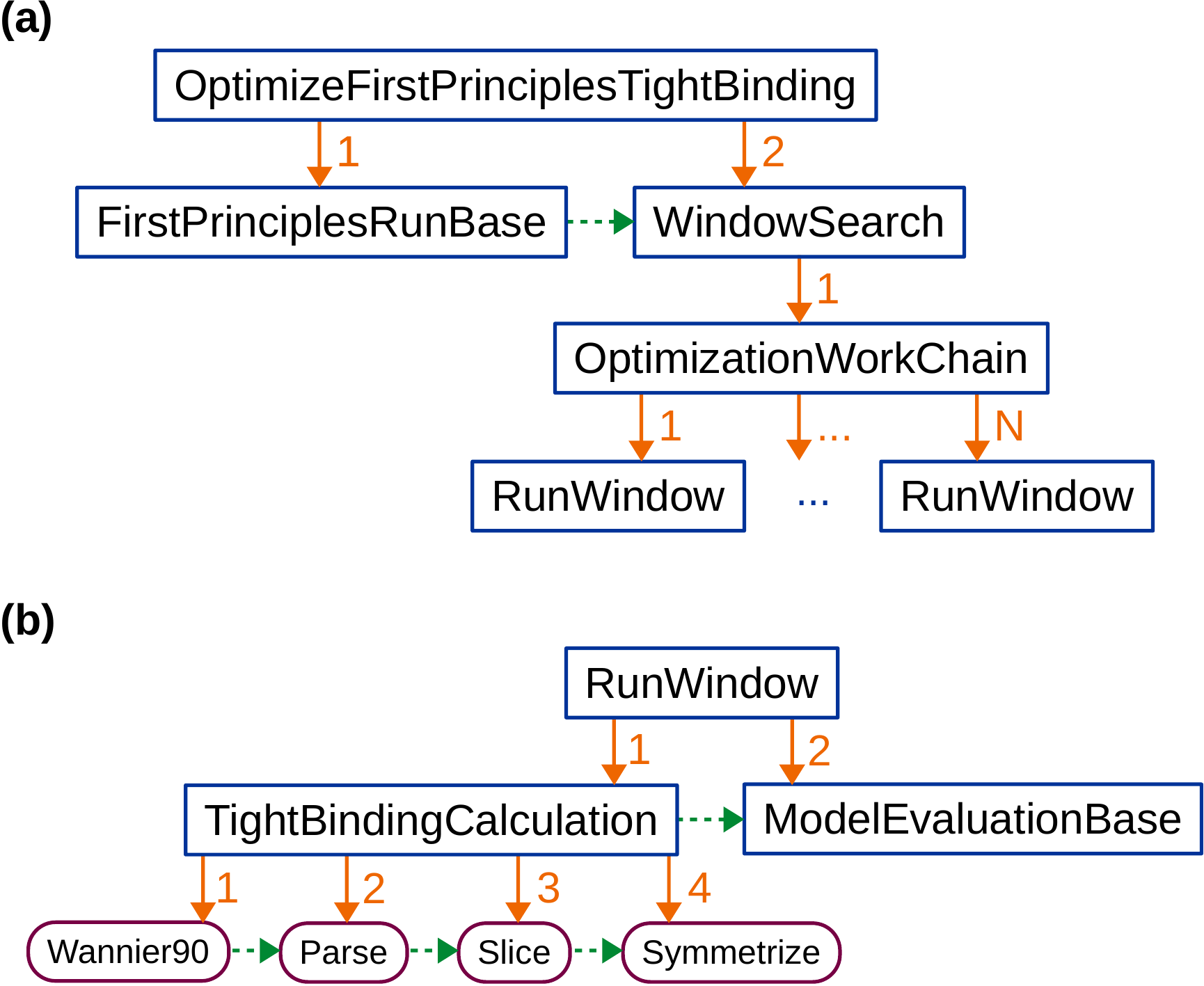}
\caption{Schematic of the AiiDA workflow for creating tight-binding models with energy window optimization. Workflows are shown in blue, and calculations in purple. Orange arrows show calls from parent- to child-workflows (or calculations). Dashed green arrows show the implicit data dependency between workflows of the same level. In calculation names, the suffix \texttt{Calculation} is omitted for brevity.}
\label{fig:workflow_diagram}
\end{figure}

Having discussed the design principles for modular workflows, we now show how these are applied to create a workflow for the construction of tight-binding models.
This workflow is implemented in the \texttt{Optimize\-First\-Principles\-Tight\-Binding} class as sketched in \cref{fig:workflow_diagram}. At the uppermost level, the workflow has two parts: \texttt{First\-Principles\-Run\-Base}, which executes the first-principles calculations, and \texttt{Window\-Search} which calculates the tight-binding model with energy window optimization.

Since different first-principles codes can produce the input files required by Wannier90, \texttt{First\-Principles\-Run\-Base} defines only the minimum interface needed to perform this task. As described in the previous section, a workflow that implements this interface for a specific first-principles code can then be chosen dynamically. As a result, the subsequent parts of the workflow are independent of which first-principles code is used.

The \texttt{Window\-Search} workflow performs the Nelder-Mead algorithm for finding the optimal energy window. Because optimization schemes are useful outside of this specific application, we implemented the Nelder-Mead method in a general way. The \texttt{Optimization\-Work\-Chain}, defined in the \texttt{aiida-optimize} module, can be used to solve generic optimization problems in the context of AiiDA workflows. It requires two inputs: A workflow which defines the function to be optimized, and an engine that implements the optimization method. Consequently, changing the whole workflow to use a different optimization method would be a simple matter of using a different engine.

Because AiiDA workflows need to be able to stop and re-start after any given step, the engine is written in an object-oriented instead of a procedural way. While this complicates implementing the Nelder-Mead method, it allows for serializing and storing the state of the engine.

The function which is optimized by the \texttt{Optimization\-Work\-Chain} is implemented in the \texttt{Run\-Window} workflow. It again consists of two parts: \texttt{Tight\-Binding\-Calculation} creates the tight-binding model itself, and \texttt{Model\-Evaluation\-Base} evaluates the quality of the model. The first step in the \texttt{Tight\-Binding\-Calculation} workflow is to run Wannier90 on the given input parameters. In a second step, the Wannier90 output is parsed and converted into the TBmodels~\cite{greschTBmodelsDocumentation} HDF5 format. A third, optional, ``slicing'' step is used to either permute the basis orbitals or discard some orbitals. Finally, the (also optional) symmetrization procedure is performed. Both the \texttt{Slice} and the \texttt{Symmetrize} calculation have a TBmodels HDF5 file as both input and output, meaning that they could be chained arbitrarily with other such post-processing steps.

For the evaluation of the tight-binding model, we again use an abstract interface class, \texttt{Model\-Evaluation\-Base}. While for the purposes of this paper we used the average difference of band energies (\cref{eqn:average_band_difference}) as a measure of model quality, other quantities might be more appropriate for different applications.

\section{Strain-dependent tight-binding models for Majorana devices}\label{sec:strain_models}

The quest for Majorana zero modes (MZMs) in condensed matter systems has recently attracted a lot of interest~\cite{kitaevUnpairedMajoranaFermions2001,lutchynMajoranaFermionsTopological2010, oregHelicalLiquidsMajorana2010,aliceaMajoranaFermionsTunable2010,pikulinZerovoltageConductancePeak2012a,aliceaNewDirectionsPursuit2012,miProposalDetectionBraiding2013,beenakkerSearchMajoranaFermions2013}. The non-abelian exchange statistics of Majorana Fermions makes these zero modes promising candidates for the realization of topological quantum computation devices~\cite{kitaevUnpairedMajoranaFermions2001,kitaevFaulttolerantQuantumComputation2003}. Experimental investigations of possible MZMs focus on the proposal by Lutchyn et. al. and Oreg et. al.~\cite{lutchynMajoranaFermionsTopological2010,oregHelicalLiquidsMajorana2010} in which MZMs appear on the boundaries of proximitized spin-orbit coupled quantum wires. Current experimental setups include semiconducting InAs nanowires with epitaxial superconducting Al~\cite{dengMajoranaBoundState2016}, and InAs/GaSb heterostructures in which the quantum spin Hall effect~\cite{kaneQuantumSpinHall2005,liuQuantumSpinHall2008} can be realized providing the possibility to proximity couple the helical edge state~\cite{pikulinZerovoltageConductancePeak2012a,miProposalDetectionBraiding2013}. While there is a good deal of evidence suggesting that MZMs exist in the wire-based setups~\cite{mourikSignaturesMajoranaFermions2012,churchillSuperconductornanowireDevicesTunneling2013}, a conclusive proof requires directly showing the braiding statistics of MZMs. 
An important step in realizing braiding with the systems based on the helical edge state is the search for optimized device and material properties. For optimizing the topological gap, a better theoretical understanding of the electronic structure in such devices is required. In this section, we show how the workflows can be used to generate tight-binding models which form the basis for accurate device simulations. While these device simulations themselves are outside the scope of this work, this shows the potential use of the method for a topic of active research in current condensed matter physics.

Highly accurate first-principles methods, using hybrid functionals~\cite{hybrids}, or the \textit{GW} approximation~\cite{gw}, are computationally too demanding for the simulation of realistic device geometries and heterostructures. State of the art simulations of such structures use the $\mathbf{k.p}$ method~\cite{kaneChapterMethod1966}, or empirical tight-binding (ETB) methods~\cite{slaterSimplifiedLCAOMethod1954}. In both of these methods the Hamiltonian is parametrized by a small number of parameters which are obtained empirically, for example via fitting to the first-principles band structure. For both of these methods the choice of parameters is ambiguous and one can obtain a good fit of the bandstructure while at the same time the electronic wavefunction might be wrongly represented. This might lead to unphysical solutions in confined geometries~\cite{tanEmpiricalTightBinding2013,tanTightbindingAnalysisSi2015}, and low transferability of the bulk models to the heterostructure in general. Recently, it was shown that better matching the ETB with the first-principles calculations can improve their transferability~\cite{tanTightbindingAnalysisSi2015,tanTransferableTightbindingModel2016}.

Realistic simulations of heterostructures require a correct treatment of strains at interfaces. In the $\mathbf{k.p}$ and the ETB method this is usually done by strain-dependent parameter sets. However, often the symmetries are not broken correctly. In this context, the Wannier or Wannier-like tight-binding models can offer a significant improvement by accurately representing the first-principles wavefunction and correctly capturing the effect of strain. As a demonstration of the AiiDA workflows, we construct SWTB models for the III-V semiconductors InSb, InAs and GaSb.

\begin{figure*}\centering
\includegraphics[width=\textwidth]{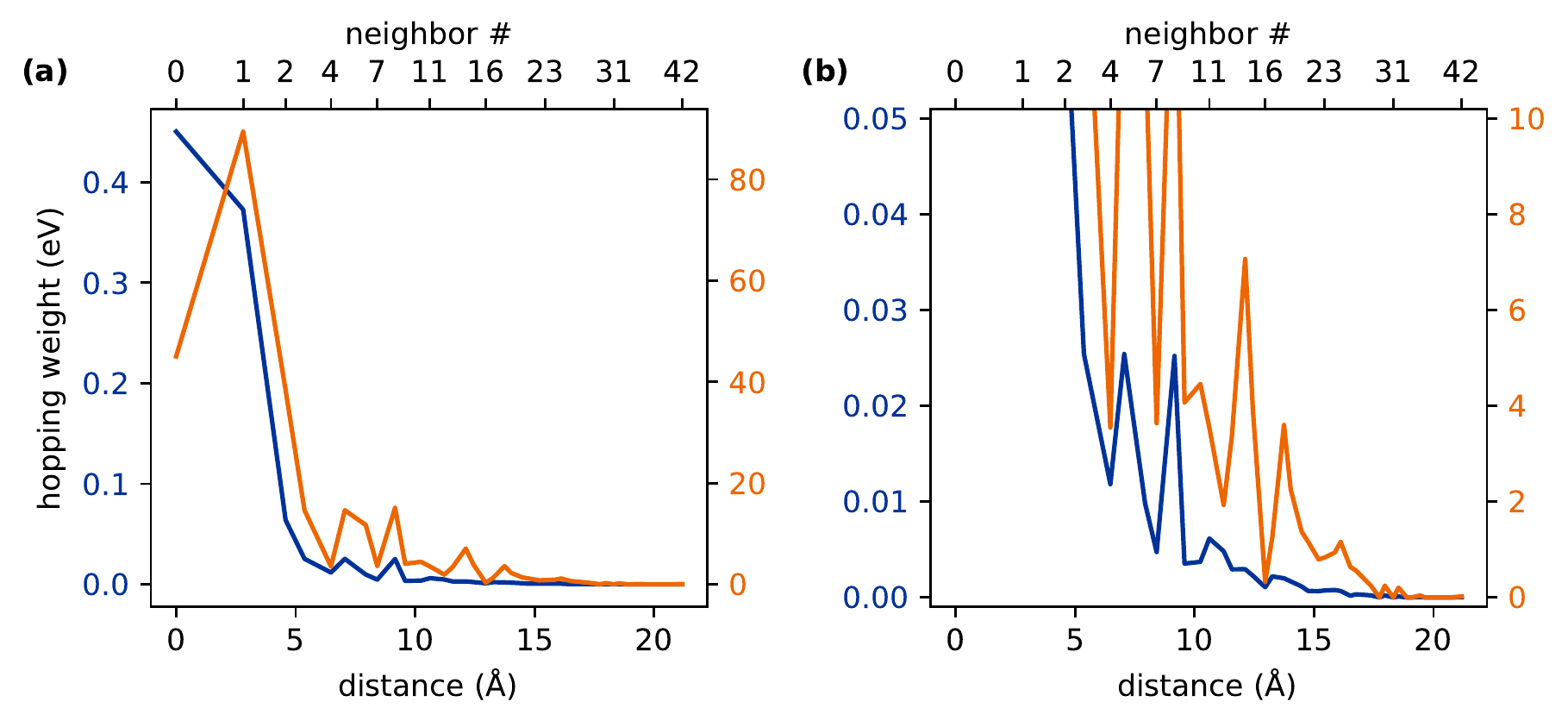}
\caption{Average (blue, left axis) and total (orange, right axis) weights of the hopping parameters for the unstrained InSb tight-binding model, as a function of distance.}
\label{fig:hopping_weights}
\end{figure*}

Including spin-orbit coupling (SOC), we require only 14 basis functions, namely $s$ and $p$ orbitals centered on the In/Ga atom, and $p$ orbitals centered on the As/Sb atom. The popular $sp^3d^5s^*$ ETB models on the other hand require 40~\cite{jancuEmpiricalMathrmspdsTightbinding1998} basis functions. The reason for this is that WTB models generally include longer-range neighbor interactions, whereas ETB is typically limited to nearest-neighbor (or next-nearest-neighbor in some cases~\cite{boykinImprovedFitsEffective1997}) interactions to keep the number of parameters manageable. As illustrated in \cref{fig:hopping_weights}, the produced tight-binding models include long-range hopping parameters, with amplitudes quickly decaying with distance.

To account for strain, we construct tight-binding models with biaxial (001), (110) and (111) strains, and the uniaxial [110] strain, as described in \cref{app:strain_details}. For each material and strain direction, we calculated $16$ models in the range of $\pm 4 \%$ strain. Including the unstrained models, we constructed a total of $195$ tight-binding models, showing the applicability of the AiiDA workflow to a large number of chemically and structurally similar compounds.

\subsection{Strained tight-binding workflow}

\begin{figure}\centering
\vspace{1em}
\includegraphics[width=\columnwidth]{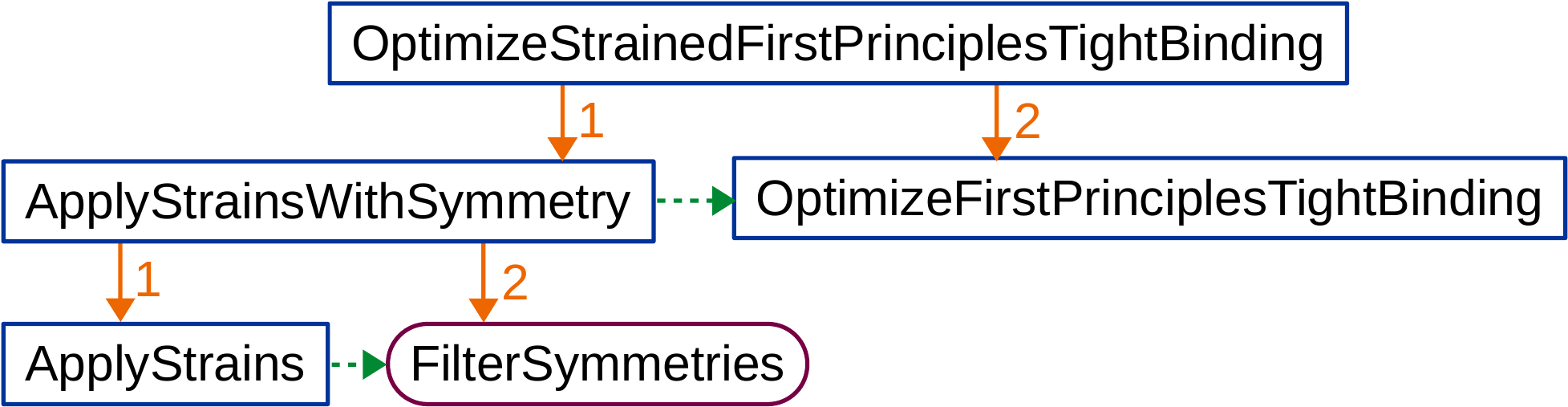}
\caption{Sketch of the workflow for constructing strained tight-binding models. The color scheme is the same as in \cref{fig:workflow_diagram}.}
\label{fig:strain_workflow_diagram}
\end{figure}

To automatically extract tight-binding models for different strain directions and strengths, we define an additional workflow, \texttt{Optimize\-Strained\-First\-Principles\-Tight\-Binding}, as shown in \cref{fig:strain_workflow_diagram}. The first step in this workflow, \texttt{Apply\-Strains\-With\-Symmetry}, creates the strained structures from the initial structure and strain parameters. Since strain can break crystal symmetries, the symmetries of the unstrained system are tested against the strained structure. With the strained structures and the remaining symmetries, we then use the \texttt{Optimize\-First\-Principles\-Tight\-Binding} workflow to create a tight-binding model for each strain value.

\subsection{First-principles calculations}

In the first step of generating the SWTB we need to carry out a first-principles calculation of the bulk semiconductor structure.
We performed all first-principles calculations using the Vienna Ab-initio Simulation Package (VASP) utilizing projector augmented-wave (PAW) basis sets~\cite{vasp}. To obtain an accurate prediction of the band gap we employed hybrid functionals~\cite{hse}. The HSE03/HSE06 hybrid functionals proved to be successful in computing band structures of III-V semiconductors~\cite{heydEfficientHybridDensity2004}. These hybrid functionals are constructed by replacing a quarter of the density functional short-range exchange (which is the Perdew-Burke-Enzerhof functional in our case~\cite{pbe}) with its Hartree-Fock counterpart. The screening parameter $\mu$ defines the separation into long- and short-range parts. In the popular HSE06 scheme, it is set to $\mu = 0.2\,\mathrm{\AA}^{-1}$. We treated $\mu$ as an empirical parameter such that the calculated band gap is fitted to the experimental value. In this work, we used $\mu_\mathrm{InAs} = 0.20\,\mathrm{\AA}^{-1}$, $\mu_\mathrm{GaSb} = 0.15\,\mathrm{\AA}^{-1}$ and $\mu_\mathrm{InSb} = 0.23\,\mathrm{\AA}^{-1}$, following the prescriptions of Ref.~\cite{kimEfficientBandStructure2010}. Since the SOC of III-V semiconductors is significant, we accounted for it by using scalar-relativistic PAW potentials.

InAs, GaSb and InSb crystallize in the zincblende structure with space group $T_d^2$ (no. 216). For the unstrained structures we perform the first-principles calculation with the experimental lattice constant $a$ at 300K, that is $a_\mathrm{InAs} = 6.058\, \mathrm{\AA}$, $a_\mathrm{GaSb} = 6.096\, \mathrm{\AA}$, $a_\mathrm{InSb} = 6.479\, \mathrm{\AA}$, from ref.~\cite{madelungCondensedMatterGroup2002}.
A plane-wave energy cutoff of $380~\mathrm{eV}$ was used for all calculations. The Brillouin-zone integrations were sampled by a $6\times6\times6$ $\Gamma$-centered $k$-points mesh.

To get optimal results from the Wannier90 code in conjunction with VASP~\cite{vasp} we found that it is necessary to turn symmetries off in VASP, that is setting the \texttt{ISYM}-tag to 0. Since the states are obtained by a numerical diagonalization routine, they obtain a random phase at each $\mathbf{k}$-point. When symmetries are enabled however, the phases are the same for all vectors forming the star of $\mathbf{k}$. Since the convergence of Wannier90 is better if the numerical phases are random, turning symmetries off generally results in more localized Wannier functions after the projection step.

\begin{figure}\centering
\includegraphics[width=0.64321608\columnwidth]{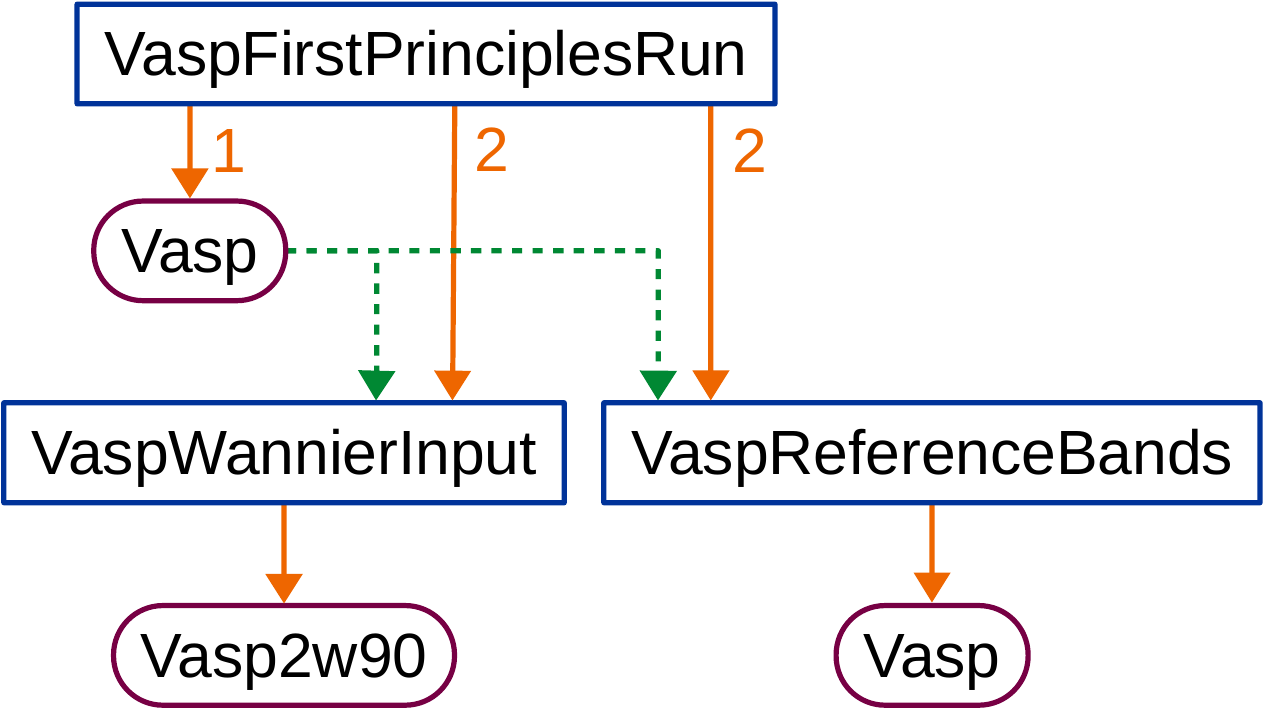}
\caption{Sketch of the \texttt{First\-Principles\-Run\-Base} subclass used for calculating the Wannier90 input and reference bands with VASP and hybrid functionals.}
\label{fig:vasp_workflow}
\end{figure}

The interface for running first-principles calculations in the tight-binding extraction workflow is defined in the \texttt{First\-Principles\-Run\-Base} class (see \cref{sec:tb_workflow}). Here, we describe the specific sub-class used to implement these calculations with VASP~\cite{vasp}, \texttt{Vasp\-First\-Principles\-Run} (see \cref{fig:vasp_workflow}). In a first step, this workflow performs a self-consistent calculation. The resulting wave-function is then passed to calculations for the reference band-structure and the input files for Wannier90. Two workflows \texttt{Vasp\-Reference\-Bands} and \texttt{Vasp\-Wannier\-Input} are used to perform these calculations. The workflows are thin wrappers around the corresponding calculations from the \texttt{aiida-vasp} plugin~\cite{hauselmannAiiDAVASPDocumentation}, providing additional input and output validation. For the band-structure calculation, the workflow also adds the $\kvec$-point grid needed for hybrid functional calculations.

\subsection{Strain interpolation}\label{sec:strain_interpolation}

Using the AiiDA workflow, we obtained tight-binding models for strains in the range of $\pm 4 \%$, in steps of $0.5 \%$. However, it is sometimes useful to have a finer control over the strain value without having to run additional first-principles calculations. A common way of obtaining this is by linear interpolation of the hopping parameters. Given two strain values $s_1$ and $s_2$, for which the hopping parameter $H^{s_i}[\Rvec]$ are known, the hopping parameters for an unknown $s^*$ can be calculated as
\begin{equation}\label{eqn:linear_interpolation}
H^{s^*}[\Rvec] = \alpha H^{s_1}[\Rvec] + (1 - \alpha) H^{s_2}[\Rvec],
\end{equation}
where
\begin{equation}
\alpha = \frac{s^* - s_2}{s_1 - s_2}.
\end{equation}

\begin{figure*}\centering
\includegraphics[width=\textwidth]{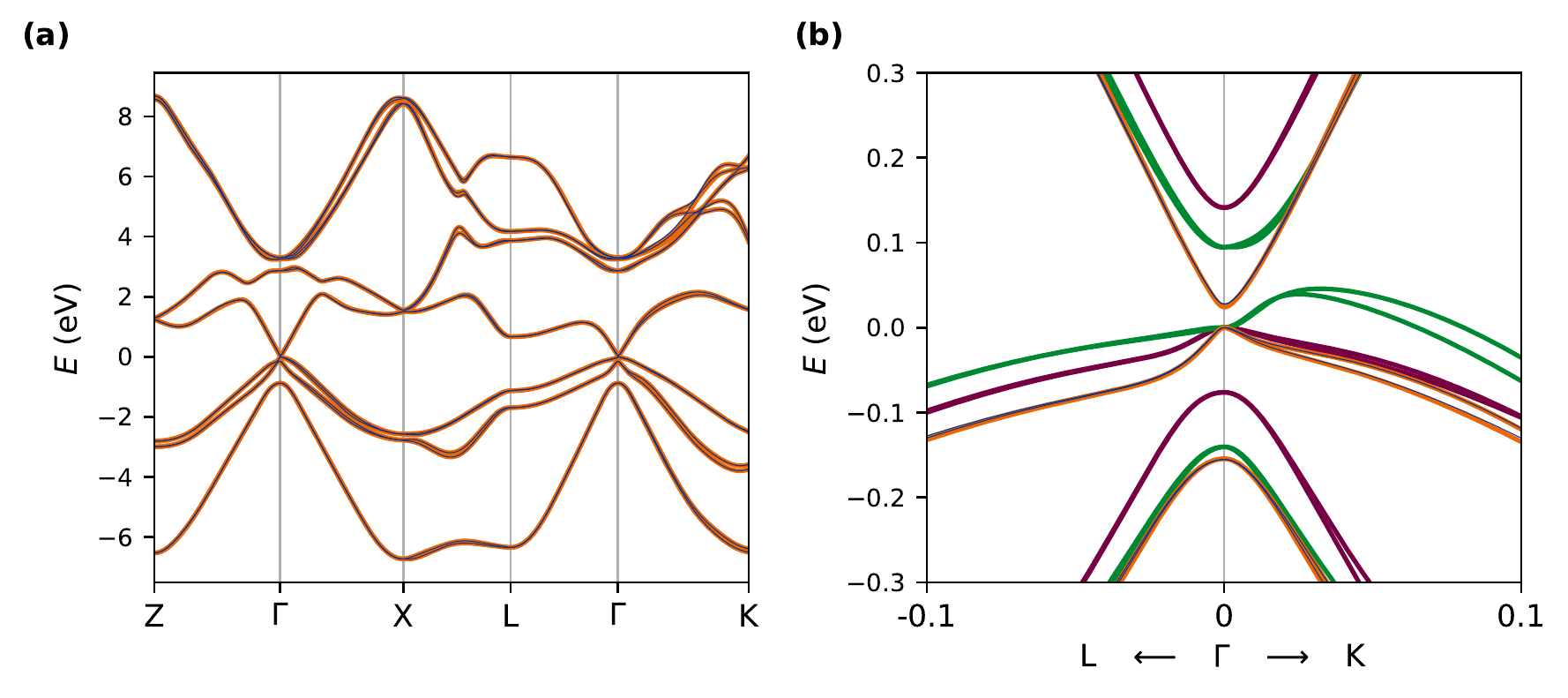}
\caption{ Comparison between the InSb bandstructure obtained directly from the tight-binding model with $2 \%$ biaxial (001) strain (blue), and from the linear interpolation (orange) between models with $1 \%$ and $3 \%$ strain. The energy scale is fixed by setting the top of the valence bands at $\Gamma$ to zero. (a) At the electron-volt scale, the only visible difference is in the upper bands along the $\Gamma$ - K line. (b) Close-up of the bands around $\Gamma$. The bands for $1 \%$ (purple) and $3 \%$ strain (green) are also shown.}
\label{fig:strain_interpolation}
\end{figure*}

Since this method assumes that the hopping parameters are a linear function of strain value, it becomes unreliable when $s^*$ is too far away from $s_1$ and $s_2$. For this reason, we compared a tight-binding model for InSb with $2 \%$ biaxial (001) strain obtained from linear interpolation of $1 \%$ and $3 \%$ strain models with one calculated directly from first-principles. \Cref{fig:strain_interpolation} shows a comparison of the two band-structures, which we find to be almost identical.

Important to note is that while linear interpolation works well for strains of the same kind, this is not necessarily the case when combining two models with different strain directions. The reason for this is that the symmetries of a particular structure depend on the direction of the applied strain, but (unless it is zero) not on its strength. As a result, a tight-binding model resulting from linear interpolation between two models of a different strain direction would not have the correct symmetries.

\subsection{Results}

To validate the tight-binding models obtained using the \texttt{aiida-tbextraction} workflows, several material parameters were calculated. \Cref{tab:eff_mass} shows effective masses and g-factors for the unstrained models, in comparison to first-principles~\cite{kimEfficientBandStructure2010} and experimental~\cite{kimEfficientBandStructure2010,vurgaftmanBandParametersIII2001} values. 
Effective masses for the tight-binding models were calculated using a second-order polynomial fit with range $0.001~\angstrom^{-1}$. The g-factor calculations were performed using both perturbation theory and a Landau level calculation~\cite{grafElectromagneticFieldsDielectric1995}, with good agreement ($< 0.5 \%$ difference) between the two methods.

\begin{table}
\begin{tabular}{@{}lllllll@{}}
\toprule
Material & Method & $|m^*_{\text{SO}}|$ & $|m^*_{\text{LH}}|$ & $|m^*_{\text{HH}}|$ & $|m^*_{\text{e}}|$ & g-factor \\
\midrule
              & $\text{HSE}_\text{bgfit}$ & 0.129~ & 0.018~ & 0.245~ & 0.017~ & \\
\textbf{InSb} & SWTB                      & 0.118 & 0.016 & 0.219 & 0.015 & -49.8 \\
              & Expt.                     & 0.110 & 0.015 & 0.263 & 0.014 & -50.6 \\
\midrule
              & $\text{HSE}_\text{bgfit}$ & 0.112 & 0.033 & 0.343 & 0.027 & \\
\textbf{InAs} & SWTB                      & 0.118 & 0.036 & 0.340 & 0.029 & -15.3 \\
              & Expt.                     & 0.140 & 0.027 & 0.333 & 0.026 & -15 \\
\midrule
              & $\text{HSE}_\text{bgfit}$ & 0.143 & 0.047 & 0.235 & 0.042 & \\
\textbf{GaSb} & SWTB                      & 0.124 & 0.039 & 0.20  & 0.036 & -15.1 \\
              & Expt.                     & 0.120 & 0.044 & 0.250 & 0.039 & -7.8 \\
\bottomrule
\end{tabular}
\caption{Effective masses of light hole (LH), heavy hole (HH), split-off hole and electron  at $\Gamma$ point along [100] direction in the unstrained case. Values for symmetrized Wannier-like tight-binding models (SWTB) are compared to first-principles (HSE$_\text{bgfit}$)~\cite{kimEfficientBandStructure2010} and experimental results~\cite{kimEfficientBandStructure2010,vurgaftmanBandParametersIII2001}.}
\label{tab:eff_mass}
\end{table}

The effect of the energy window optimization is shown in \cref{tab:energy_window_opt}, which lists the initial and optimized windows, as well as the corresponding band-structure mismatch. As previously shown in \cref{fig:optimize_energy_window}, it can be seen that the mismatch is substantially reduced after optimization.

\begin{table}
\begin{tabular}{@{}lllllll@{}}
\toprule
\multicolumn{2}{@{}l}{\textbf{Material}} & \multicolumn{4}{l}{\textbf{Energy Windows (eV)}} & \textbf{$\Delta$}\\
\midrule
\multirow{2}{*}{InSb~} & initial   & $(-4.5,$ & $[-4,$ & $6.5],$ & $16)$ & $0.107$ \\
                      & optimized~ & $(-4.44,$ & $[-3.24,$ & $8.67],$ & $14.01)$~ & $0.033$ \\
\midrule
\multirow{2}{*}{InAs~} & initial   & $(-4.5,$ & $[-4,$ & $6.5],$ & $16)$ & $0.113$ \\
                      & optimized & $(-4.44,$ & $[-3.59,$ & $7.34],$ & $15.04)$	& $0.046$ \\
\midrule
\multirow{2}{*}{GaSb~} & initial   & $(-4.5,$ & $[-4.5,$ & $7],$ & $16)$ & $0.082$ \\
                      & optimized & $(-5.35,$ & $[-3.34,$ & $7.90],$ & $14.27)$ & $0.043$ \\
\bottomrule
\end{tabular}
\caption{Initial and optimized energy windows used for calculating unstrained tight-binding models, and the corresponding band-structure mismatch as defined in \cref{eqn:average_band_difference}.}
\label{tab:energy_window_opt}
\end{table}

Finally, the effect of strain on the energy levels at high-symmetry points is shown in \cref{fig:strain_band_shift}. The numerical data is listed in the supplementary files~\footnote{See Supplemental Material at [URL] for tables containing the band energies at high-symmetry points for different values of strain.}.

\begin{figure*}\centering
\includegraphics[width=\textwidth]{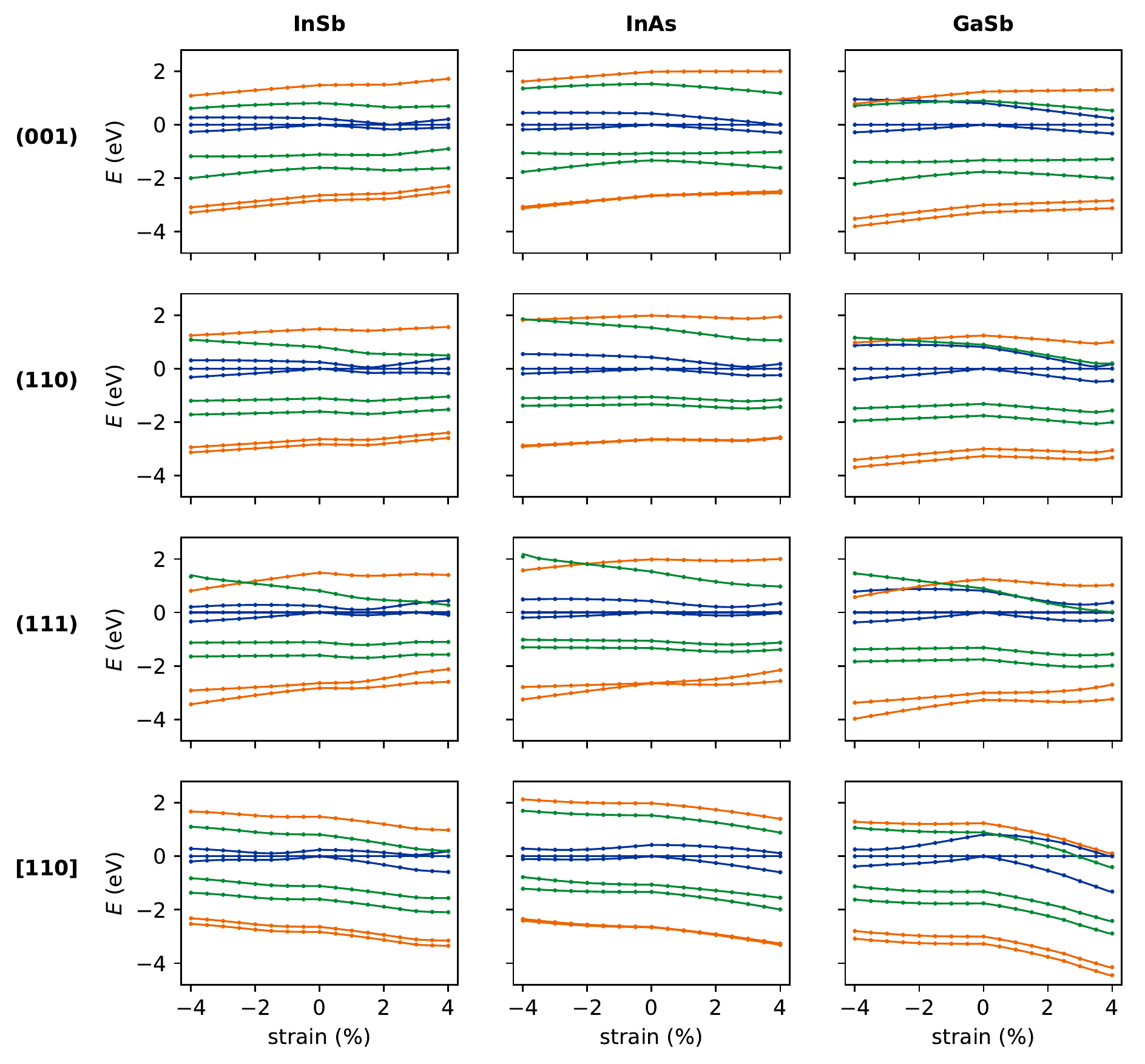}
\caption{Strain dependence of band energies. The two highest valence bands and the lowest conduction band are shown at the $\Gamma$ (blue), $X$ (orange) and $L$ (green) points, where each band is doubly degenerate. Energy values are shifted such that the valence band maximum at $\Gamma$ is zero. The line represents values calculated from the tight-binding models with linear interpolation (\cref{eqn:linear_interpolation}) in steps of $0.1 \%$. For comparison, the points show values calculated from first-principles. We find a good agreement between the tight-binding and first-principles values, except for the conduction band value at the $L$ - point at $-4\%$ biaxial $(111)$ strain.}
\label{fig:strain_band_shift}
\end{figure*}

In the supplementary materials of this paper, an export of the AiiDA database is given~\footnote{See Supplemental Material at [URL] for an full export of the AiiDA database containing the calculations of strained InSb, InAs, and GaSb tight-binding models.}. This database contains the full provenance of each calculation performed to create the tight-binding models. For ease of accessibility, a separate data set containing only the $195$ strained tight-binding models is also given~\footnote{See Supplemental Material at [URL] for an archive containing the $195$ strained tight-binding models of InSb, InAs, and GaSb.}.

\section{Conclusion and Outlook}
\label{sec4}

We have implemented a workflow for an \textit{automatic} construction of Wannier tight-binding models from first-principles calculations. Building on the known procedure for calculating these models, we introduced a post-processing step to symmetrize the models, and an optimization of the energy windows used for disentanglement. These workflows are implemented in the \texttt{aiida-tbextraction} package, which is a free and open-source plugin for the AiiDA framework. As a test case, tight-binding models for strained III-V semiconductor materials were calculated. These results should enable device simulations for Majorana designs and other quantum devices.

The workflows have been implemented in a modular and extensible way. As a result, they can be used as building blocks for further improvements in automating the process of generating Wannier tight-binding models. Possible directions include extending the number of first-principles codes which are compatible with the plugin, adding different fitness criteria for the energy window optimization, and further minimizing the number of tunable parameters. For example, the need for choosing initial trial orbitals could be eliminated either by using another optimization step, or by utilizing the method of Ref.~\cite{mustafaAutomatedConstructionMaximally2015}.

\section*{Acknowledgments}
We would like to thank S. Huber, M. Uhrin, G. Pizzi, A. Marrazzo, M. K\"onz and D. Rodic for helpful discussions. The authors were supported by Microsoft Research, and the Swiss National Science Foundation through the National  Competence  Centers in Research MARVEL and QSIT. AAS also acknowledges the support of the Swiss National Science Foundation Professorship. Calculations were performed on the M\"onch cluster of ETH Zurich.

\bibliography{refs_zotero,refs_manual}

\begin{thebibliography}{68}%
\makeatletter
\providecommand \@ifxundefined [1]{%
 \@ifx{#1\undefined}
}%
\providecommand \@ifnum [1]{%
 \ifnum #1\expandafter \@firstoftwo
 \else \expandafter \@secondoftwo
 \fi
}%
\providecommand \@ifx [1]{%
 \ifx #1\expandafter \@firstoftwo
 \else \expandafter \@secondoftwo
 \fi
}%
\providecommand \natexlab [1]{#1}%
\providecommand \enquote  [1]{``#1''}%
\providecommand \bibnamefont  [1]{#1}%
\providecommand \bibfnamefont [1]{#1}%
\providecommand \citenamefont [1]{#1}%
\providecommand \href@noop [0]{\@secondoftwo}%
\providecommand \href [0]{\begingroup \@sanitize@url \@href}%
\providecommand \@href[1]{\@@startlink{#1}\@@href}%
\providecommand \@@href[1]{\endgroup#1\@@endlink}%
\providecommand \@sanitize@url [0]{\catcode `\\12\catcode `\$12\catcode
  `\&12\catcode `\#12\catcode `\^12\catcode `\_12\catcode `\%12\relax}%
\providecommand \@@startlink[1]{}%
\providecommand \@@endlink[0]{}%
\providecommand \url  [0]{\begingroup\@sanitize@url \@url }%
\providecommand \@url [1]{\endgroup\@href {#1}{\urlprefix }}%
\providecommand \urlprefix  [0]{URL }%
\providecommand \Eprint [0]{\href }%
\providecommand \doibase [0]{http://dx.doi.org/}%
\providecommand \selectlanguage [0]{\@gobble}%
\providecommand \bibinfo  [0]{\@secondoftwo}%
\providecommand \bibfield  [0]{\@secondoftwo}%
\providecommand \translation [1]{[#1]}%
\providecommand \BibitemOpen [0]{}%
\providecommand \bibitemStop [0]{}%
\providecommand \bibitemNoStop [0]{.\EOS\space}%
\providecommand \EOS [0]{\spacefactor3000\relax}%
\providecommand \BibitemShut  [1]{\csname bibitem#1\endcsname}%
\let\auto@bib@innerbib\@empty
\bibitem [{\citenamefont {Hohenberg}\ and\ \citenamefont
  {Kohn}(1964)}]{hohenberg_kohn_dft}%
  \BibitemOpen
  \bibfield  {author} {\bibinfo {author} {\bibfnamefont {P.}~\bibnamefont
  {Hohenberg}}\ and\ \bibinfo {author} {\bibfnamefont {W.}~\bibnamefont
  {Kohn}},\ }\href {\doibase 10.1103/PhysRev.136.B864} {\bibfield  {journal}
  {\bibinfo  {journal} {Phys. Rev.}\ }\textbf {\bibinfo {volume} {136}},\
  \bibinfo {pages} {B864} (\bibinfo {year} {1964})}\BibitemShut {NoStop}%
\bibitem [{\citenamefont {Marzari}\ and\ \citenamefont
  {Vanderbilt}(1997)}]{marzariMaximallyLocalizedGeneralized1997}%
  \BibitemOpen
  \bibfield  {author} {\bibinfo {author} {\bibfnamefont {N.}~\bibnamefont
  {Marzari}}\ and\ \bibinfo {author} {\bibfnamefont {D.}~\bibnamefont
  {Vanderbilt}},\ }\href {\doibase 10.1103/PhysRevB.56.12847} {\bibfield
  {journal} {\bibinfo  {journal} {Phys. Rev. B}\ }\textbf {\bibinfo {volume}
  {56}},\ \bibinfo {pages} {12847} (\bibinfo {year} {1997})}\BibitemShut
  {NoStop}%
\bibitem [{\citenamefont {Souza}\ \emph {et~al.}(2001)\citenamefont {Souza},
  \citenamefont {Marzari},\ and\ \citenamefont
  {Vanderbilt}}]{souzaMaximallyLocalizedWannier2001}%
  \BibitemOpen
  \bibfield  {author} {\bibinfo {author} {\bibfnamefont {I.}~\bibnamefont
  {Souza}}, \bibinfo {author} {\bibfnamefont {N.}~\bibnamefont {Marzari}}, \
  and\ \bibinfo {author} {\bibfnamefont {D.}~\bibnamefont {Vanderbilt}},\
  }\href {\doibase 10.1103/PhysRevB.65.035109} {\bibfield  {journal} {\bibinfo
  {journal} {Phys. Rev. B}\ }\textbf {\bibinfo {volume} {65}},\ \bibinfo
  {pages} {035109} (\bibinfo {year} {2001})}\BibitemShut {NoStop}%
\bibitem [{\citenamefont {Franceschetti}\ and\ \citenamefont
  {Zunger}(1999)}]{franceschettiInverseBandstructureProblem1999}%
  \BibitemOpen
  \bibfield  {author} {\bibinfo {author} {\bibfnamefont {A.}~\bibnamefont
  {Franceschetti}}\ and\ \bibinfo {author} {\bibfnamefont {A.}~\bibnamefont
  {Zunger}},\ }\href {\doibase 10.1038/46995} {\bibfield  {journal} {\bibinfo
  {journal} {Nature}\ }\textbf {\bibinfo {volume} {402}},\ \bibinfo {pages}
  {60} (\bibinfo {year} {1999})}\BibitemShut {NoStop}%
\bibitem [{\citenamefont {J\'ohannesson}\ \emph {et~al.}(2002)\citenamefont
  {J\'ohannesson}, \citenamefont {Bligaard}, \citenamefont {Ruban},
  \citenamefont {Skriver}, \citenamefont {Jacobsen},\ and\ \citenamefont
  {N\o{}rskov}}]{johannessonCombinedElectronicStructure2002}%
  \BibitemOpen
  \bibfield  {author} {\bibinfo {author} {\bibfnamefont {G.~H.}\ \bibnamefont
  {J\'ohannesson}}, \bibinfo {author} {\bibfnamefont {T.}~\bibnamefont
  {Bligaard}}, \bibinfo {author} {\bibfnamefont {A.~V.}\ \bibnamefont {Ruban}},
  \bibinfo {author} {\bibfnamefont {H.~L.}\ \bibnamefont {Skriver}}, \bibinfo
  {author} {\bibfnamefont {K.~W.}\ \bibnamefont {Jacobsen}}, \ and\ \bibinfo
  {author} {\bibfnamefont {J.~K.}\ \bibnamefont {N\o{}rskov}},\ }\href
  {\doibase 10.1103/PhysRevLett.88.255506} {\bibfield  {journal} {\bibinfo
  {journal} {Phys. Rev. Lett.}\ }\textbf {\bibinfo {volume} {88}},\ \bibinfo
  {pages} {255506} (\bibinfo {year} {2002})}\BibitemShut {NoStop}%
\bibitem [{\citenamefont {Curtarolo}\ \emph {et~al.}(2003)\citenamefont
  {Curtarolo}, \citenamefont {Morgan}, \citenamefont {Persson}, \citenamefont
  {Rodgers},\ and\ \citenamefont
  {Ceder}}]{curtaroloPredictingCrystalStructures2003}%
  \BibitemOpen
  \bibfield  {author} {\bibinfo {author} {\bibfnamefont {S.}~\bibnamefont
  {Curtarolo}}, \bibinfo {author} {\bibfnamefont {D.}~\bibnamefont {Morgan}},
  \bibinfo {author} {\bibfnamefont {K.}~\bibnamefont {Persson}}, \bibinfo
  {author} {\bibfnamefont {J.}~\bibnamefont {Rodgers}}, \ and\ \bibinfo
  {author} {\bibfnamefont {G.}~\bibnamefont {Ceder}},\ }\href {\doibase
  10.1103/PhysRevLett.91.135503} {\bibfield  {journal} {\bibinfo  {journal}
  {Phys. Rev. Lett.}\ }\textbf {\bibinfo {volume} {91}},\ \bibinfo {pages}
  {135503} (\bibinfo {year} {2003})}\BibitemShut {NoStop}%
\bibitem [{\citenamefont {Curtarolo}\ \emph {et~al.}(2013)\citenamefont
  {Curtarolo}, \citenamefont {Hart}, \citenamefont {Nardelli}, \citenamefont
  {Mingo}, \citenamefont {Sanvito},\ and\ \citenamefont
  {Levy}}]{curtaroloHighthroughputHighwayComputational2013}%
  \BibitemOpen
  \bibfield  {author} {\bibinfo {author} {\bibfnamefont {S.}~\bibnamefont
  {Curtarolo}}, \bibinfo {author} {\bibfnamefont {G.~L.~W.}\ \bibnamefont
  {Hart}}, \bibinfo {author} {\bibfnamefont {M.~B.}\ \bibnamefont {Nardelli}},
  \bibinfo {author} {\bibfnamefont {N.}~\bibnamefont {Mingo}}, \bibinfo
  {author} {\bibfnamefont {S.}~\bibnamefont {Sanvito}}, \ and\ \bibinfo
  {author} {\bibfnamefont {O.}~\bibnamefont {Levy}},\ }\href {\doibase
  10.1038/nmat3568} {\bibfield  {journal} {\bibinfo  {journal} {Nat. Mater.}\
  }\textbf {\bibinfo {volume} {12}},\ \bibinfo {pages} {191} (\bibinfo {year}
  {2013})}\BibitemShut {NoStop}%
\bibitem [{\citenamefont {Pizzi}\ \emph {et~al.}(2016)\citenamefont {Pizzi},
  \citenamefont {Cepellotti}, \citenamefont {Sabatini}, \citenamefont
  {Marzari},\ and\ \citenamefont {Kozinsky}}]{aiida}%
  \BibitemOpen
  \bibfield  {author} {\bibinfo {author} {\bibfnamefont {G.}~\bibnamefont
  {Pizzi}}, \bibinfo {author} {\bibfnamefont {A.}~\bibnamefont {Cepellotti}},
  \bibinfo {author} {\bibfnamefont {R.}~\bibnamefont {Sabatini}}, \bibinfo
  {author} {\bibfnamefont {N.}~\bibnamefont {Marzari}}, \ and\ \bibinfo
  {author} {\bibfnamefont {B.}~\bibnamefont {Kozinsky}},\ }\href {\doibase
  10.1016/j.commatsci.2015.09.013} {\bibfield  {journal} {\bibinfo  {journal}
  {Computational Materials Science}\ }\textbf {\bibinfo {volume} {111}},\
  \bibinfo {pages} {218} (\bibinfo {year} {2016})}\BibitemShut {NoStop}%
\bibitem [{\citenamefont {Jain}\ \emph {et~al.}(2015)\citenamefont {Jain},
  \citenamefont {Ong}, \citenamefont {Chen}, \citenamefont {Medasani},
  \citenamefont {Qu}, \citenamefont {Kocher}, \citenamefont {Brafman},
  \citenamefont {Petretto}, \citenamefont {Rignanese}, \citenamefont {Hautier},
  \citenamefont {Gunter},\ and\ \citenamefont
  {Persson}}]{jainFireWorksDynamicWorkflow2015}%
  \BibitemOpen
  \bibfield  {author} {\bibinfo {author} {\bibfnamefont {A.}~\bibnamefont
  {Jain}}, \bibinfo {author} {\bibfnamefont {S.~P.}\ \bibnamefont {Ong}},
  \bibinfo {author} {\bibfnamefont {W.}~\bibnamefont {Chen}}, \bibinfo {author}
  {\bibfnamefont {B.}~\bibnamefont {Medasani}}, \bibinfo {author}
  {\bibfnamefont {X.}~\bibnamefont {Qu}}, \bibinfo {author} {\bibfnamefont
  {M.}~\bibnamefont {Kocher}}, \bibinfo {author} {\bibfnamefont
  {M.}~\bibnamefont {Brafman}}, \bibinfo {author} {\bibfnamefont
  {G.}~\bibnamefont {Petretto}}, \bibinfo {author} {\bibfnamefont {G.-M.}\
  \bibnamefont {Rignanese}}, \bibinfo {author} {\bibfnamefont {G.}~\bibnamefont
  {Hautier}}, \bibinfo {author} {\bibfnamefont {D.}~\bibnamefont {Gunter}}, \
  and\ \bibinfo {author} {\bibfnamefont {K.~A.}\ \bibnamefont {Persson}},\
  }\href {\doibase 10.1002/cpe.3505} {\bibfield  {journal} {\bibinfo  {journal}
  {Concurr. Comput. Pract. Exp.}\ }\textbf {\bibinfo {volume} {27}},\ \bibinfo
  {pages} {5037} (\bibinfo {year} {2015})}\BibitemShut {NoStop}%
\bibitem [{\citenamefont {Mostofi}\ \emph {et~al.}(2008)\citenamefont
  {Mostofi}, \citenamefont {Yates}, \citenamefont {Lee}, \citenamefont {Souza},
  \citenamefont {Vanderbilt},\ and\ \citenamefont {Marzari}}]{wannier90}%
  \BibitemOpen
  \bibfield  {author} {\bibinfo {author} {\bibfnamefont {A.~A.}\ \bibnamefont
  {Mostofi}}, \bibinfo {author} {\bibfnamefont {J.~R.}\ \bibnamefont {Yates}},
  \bibinfo {author} {\bibfnamefont {Y.-S.}\ \bibnamefont {Lee}}, \bibinfo
  {author} {\bibfnamefont {I.}~\bibnamefont {Souza}}, \bibinfo {author}
  {\bibfnamefont {D.}~\bibnamefont {Vanderbilt}}, \ and\ \bibinfo {author}
  {\bibfnamefont {N.}~\bibnamefont {Marzari}},\ }\href {\doibase
  10.1016/j.cpc.2007.11.016} {\bibfield  {journal} {\bibinfo  {journal}
  {Computer Physics Communications}\ }\textbf {\bibinfo {volume} {178}},\
  \bibinfo {pages} {685} (\bibinfo {year} {2008})}\BibitemShut {NoStop}%
\bibitem [{\citenamefont {Mostofi}\ \emph {et~al.}(2014)\citenamefont
  {Mostofi}, \citenamefont {Yates}, \citenamefont {Pizzi}, \citenamefont {Lee},
  \citenamefont {Souza}, \citenamefont {Vanderbilt},\ and\ \citenamefont
  {Marzari}}]{wannier90_updated}%
  \BibitemOpen
  \bibfield  {author} {\bibinfo {author} {\bibfnamefont {A.~A.}\ \bibnamefont
  {Mostofi}}, \bibinfo {author} {\bibfnamefont {J.~R.}\ \bibnamefont {Yates}},
  \bibinfo {author} {\bibfnamefont {G.}~\bibnamefont {Pizzi}}, \bibinfo
  {author} {\bibfnamefont {Y.-S.}\ \bibnamefont {Lee}}, \bibinfo {author}
  {\bibfnamefont {I.}~\bibnamefont {Souza}}, \bibinfo {author} {\bibfnamefont
  {D.}~\bibnamefont {Vanderbilt}}, \ and\ \bibinfo {author} {\bibfnamefont
  {N.}~\bibnamefont {Marzari}},\ }\href {\doibase 10.1016/j.cpc.2014.05.003}
  {\bibfield  {journal} {\bibinfo  {journal} {Computer Physics Communications}\
  }\textbf {\bibinfo {volume} {185}},\ \bibinfo {pages} {2309} (\bibinfo {year}
  {2014})}\BibitemShut {NoStop}%
\bibitem [{\citenamefont {Mustafa}\ \emph {et~al.}(2015)\citenamefont
  {Mustafa}, \citenamefont {Coh}, \citenamefont {Cohen},\ and\ \citenamefont
  {Louie}}]{mustafaAutomatedConstructionMaximally2015}%
  \BibitemOpen
  \bibfield  {author} {\bibinfo {author} {\bibfnamefont {J.~I.}\ \bibnamefont
  {Mustafa}}, \bibinfo {author} {\bibfnamefont {S.}~\bibnamefont {Coh}},
  \bibinfo {author} {\bibfnamefont {M.~L.}\ \bibnamefont {Cohen}}, \ and\
  \bibinfo {author} {\bibfnamefont {S.~G.}\ \bibnamefont {Louie}},\ }\href
  {\doibase 10.1103/PhysRevB.92.165134} {\bibfield  {journal} {\bibinfo
  {journal} {Phys. Rev. B}\ }\textbf {\bibinfo {volume} {92}},\ \bibinfo
  {pages} {165134} (\bibinfo {year} {2015})}\BibitemShut {NoStop}%
\bibitem [{\citenamefont {Kitaev}(2001)}]{kitaevUnpairedMajoranaFermions2001}%
  \BibitemOpen
  \bibfield  {author} {\bibinfo {author} {\bibfnamefont {A.~Y.}\ \bibnamefont
  {Kitaev}},\ }\href {\doibase 10.1070/1063-7869/44/10S/S29} {\bibfield
  {journal} {\bibinfo  {journal} {Phys.-Usp.}\ }\textbf {\bibinfo {volume}
  {44}},\ \bibinfo {pages} {131} (\bibinfo {year} {2001})}\BibitemShut
  {NoStop}%
\bibitem [{\citenamefont {Lutchyn}\ \emph {et~al.}(2010)\citenamefont
  {Lutchyn}, \citenamefont {Sau},\ and\ \citenamefont
  {Das~Sarma}}]{lutchynMajoranaFermionsTopological2010}%
  \BibitemOpen
  \bibfield  {author} {\bibinfo {author} {\bibfnamefont {R.~M.}\ \bibnamefont
  {Lutchyn}}, \bibinfo {author} {\bibfnamefont {J.~D.}\ \bibnamefont {Sau}}, \
  and\ \bibinfo {author} {\bibfnamefont {S.}~\bibnamefont {Das~Sarma}},\ }\href
  {\doibase 10.1103/PhysRevLett.105.077001} {\bibfield  {journal} {\bibinfo
  {journal} {Phys. Rev. Lett.}\ }\textbf {\bibinfo {volume} {105}},\ \bibinfo
  {pages} {077001} (\bibinfo {year} {2010})}\BibitemShut {NoStop}%
\bibitem [{\citenamefont {Oreg}\ \emph {et~al.}(2010)\citenamefont {Oreg},
  \citenamefont {Refael},\ and\ \citenamefont {{von
  Oppen}}}]{oregHelicalLiquidsMajorana2010}%
  \BibitemOpen
  \bibfield  {author} {\bibinfo {author} {\bibfnamefont {Y.}~\bibnamefont
  {Oreg}}, \bibinfo {author} {\bibfnamefont {G.}~\bibnamefont {Refael}}, \ and\
  \bibinfo {author} {\bibfnamefont {F.}~\bibnamefont {{von Oppen}}},\ }\href
  {\doibase 10.1103/PhysRevLett.105.177002} {\bibfield  {journal} {\bibinfo
  {journal} {Phys. Rev. Lett.}\ }\textbf {\bibinfo {volume} {105}},\ \bibinfo
  {pages} {177002} (\bibinfo {year} {2010})}\BibitemShut {NoStop}%
\bibitem [{Note1()}]{Note1}%
  \BibitemOpen
  \bibinfo {note} {In this work, we use the tight-binding convention I of
  Ref.~\cite {pythtb_formalism}.}\BibitemShut {Stop}%
\bibitem [{\citenamefont {Dresselhaus}\ \emph {et~al.}(2007)\citenamefont
  {Dresselhaus}, \citenamefont {Dresselhaus},\ and\ \citenamefont
  {Jorio}}]{dresselhausGroupTheoryApplication2007}%
  \BibitemOpen
  \bibfield  {author} {\bibinfo {author} {\bibfnamefont {M.~S.}\ \bibnamefont
  {Dresselhaus}}, \bibinfo {author} {\bibfnamefont {G.}~\bibnamefont
  {Dresselhaus}}, \ and\ \bibinfo {author} {\bibfnamefont {A.}~\bibnamefont
  {Jorio}},\ }\href@noop {} {{\selectlanguage {english}\emph {\bibinfo {title}
  {Group {{Theory}}: {{Application}} to the {{Physics}} of {{Condensed
  Matter}}}}}}\ (\bibinfo  {publisher} {{Springer Science \& Business Media}},\
  \bibinfo {year} {2007})\BibitemShut {NoStop}%
\bibitem [{\citenamefont
  {Haber}(2011)}]{haberThreeDimensionalProperImproper2011}%
  \BibitemOpen
  \bibfield  {author} {\bibinfo {author} {\bibfnamefont {H.~E.}\ \bibnamefont
  {Haber}},\ }\href@noop {} {\enquote {\bibinfo {title} {Three-{{Dimensional
  Proper}} and {{Improper Rotation Matrices}}},}\ } (\bibinfo {year}
  {2011})\BibitemShut {NoStop}%
\bibitem [{\citenamefont
  {Sakuma}(2013)}]{sakumaSymmetryadaptedWannierFunctions2013}%
  \BibitemOpen
  \bibfield  {author} {\bibinfo {author} {\bibfnamefont {R.}~\bibnamefont
  {Sakuma}},\ }\href {\doibase 10.1103/PhysRevB.87.235109} {\bibfield
  {journal} {\bibinfo  {journal} {Phys. Rev. B}\ }\textbf {\bibinfo {volume}
  {87}},\ \bibinfo {pages} {235109} (\bibinfo {year} {2013})}\BibitemShut
  {NoStop}%
\bibitem [{\citenamefont {Giannozzi}\ \emph {et~al.}(2009)\citenamefont
  {Giannozzi}, \citenamefont {Baroni}, \citenamefont {Bonini}, \citenamefont
  {Calandra}, \citenamefont {Car}, \citenamefont {Cavazzoni}, \citenamefont
  {{Davide Ceresoli}}, \citenamefont {Chiarotti}, \citenamefont {Cococcioni},
  \citenamefont {Dabo}, \citenamefont {Corso}, \citenamefont {de~Gironcoli},
  \citenamefont {Fabris}, \citenamefont {Fratesi}, \citenamefont {Gebauer},
  \citenamefont {Gerstmann}, \citenamefont {Gougoussis}, \citenamefont {{Anton
  Kokalj}}, \citenamefont {Lazzeri}, \citenamefont {{Martin-Samos}},
  \citenamefont {Marzari}, \citenamefont {Mauri}, \citenamefont {Mazzarello},
  \citenamefont {{Stefano Paolini}}, \citenamefont {Pasquarello}, \citenamefont
  {Paulatto}, \citenamefont {Sbraccia}, \citenamefont {Scandolo}, \citenamefont
  {Sclauzero}, \citenamefont {Seitsonen}, \citenamefont {Smogunov},
  \citenamefont {Umari},\ and\ \citenamefont {Wentzcovitch}}]{espresso}%
  \BibitemOpen
  \bibfield  {author} {\bibinfo {author} {\bibfnamefont {P.}~\bibnamefont
  {Giannozzi}}, \bibinfo {author} {\bibfnamefont {S.}~\bibnamefont {Baroni}},
  \bibinfo {author} {\bibfnamefont {N.}~\bibnamefont {Bonini}}, \bibinfo
  {author} {\bibfnamefont {M.}~\bibnamefont {Calandra}}, \bibinfo {author}
  {\bibfnamefont {R.}~\bibnamefont {Car}}, \bibinfo {author} {\bibfnamefont
  {C.}~\bibnamefont {Cavazzoni}}, \bibinfo {author} {\bibnamefont {{Davide
  Ceresoli}}}, \bibinfo {author} {\bibfnamefont {G.~L.}\ \bibnamefont
  {Chiarotti}}, \bibinfo {author} {\bibfnamefont {M.}~\bibnamefont
  {Cococcioni}}, \bibinfo {author} {\bibfnamefont {I.}~\bibnamefont {Dabo}},
  \bibinfo {author} {\bibfnamefont {A.~D.}\ \bibnamefont {Corso}}, \bibinfo
  {author} {\bibfnamefont {S.}~\bibnamefont {de~Gironcoli}}, \bibinfo {author}
  {\bibfnamefont {S.}~\bibnamefont {Fabris}}, \bibinfo {author} {\bibfnamefont
  {G.}~\bibnamefont {Fratesi}}, \bibinfo {author} {\bibfnamefont
  {R.}~\bibnamefont {Gebauer}}, \bibinfo {author} {\bibfnamefont
  {U.}~\bibnamefont {Gerstmann}}, \bibinfo {author} {\bibfnamefont
  {C.}~\bibnamefont {Gougoussis}}, \bibinfo {author} {\bibnamefont {{Anton
  Kokalj}}}, \bibinfo {author} {\bibfnamefont {M.}~\bibnamefont {Lazzeri}},
  \bibinfo {author} {\bibfnamefont {L.}~\bibnamefont {{Martin-Samos}}},
  \bibinfo {author} {\bibfnamefont {N.}~\bibnamefont {Marzari}}, \bibinfo
  {author} {\bibfnamefont {F.}~\bibnamefont {Mauri}}, \bibinfo {author}
  {\bibfnamefont {R.}~\bibnamefont {Mazzarello}}, \bibinfo {author}
  {\bibnamefont {{Stefano Paolini}}}, \bibinfo {author} {\bibfnamefont
  {A.}~\bibnamefont {Pasquarello}}, \bibinfo {author} {\bibfnamefont
  {L.}~\bibnamefont {Paulatto}}, \bibinfo {author} {\bibfnamefont
  {C.}~\bibnamefont {Sbraccia}}, \bibinfo {author} {\bibfnamefont
  {S.}~\bibnamefont {Scandolo}}, \bibinfo {author} {\bibfnamefont
  {G.}~\bibnamefont {Sclauzero}}, \bibinfo {author} {\bibfnamefont {A.~P.}\
  \bibnamefont {Seitsonen}}, \bibinfo {author} {\bibfnamefont {A.}~\bibnamefont
  {Smogunov}}, \bibinfo {author} {\bibfnamefont {P.}~\bibnamefont {Umari}}, \
  and\ \bibinfo {author} {\bibfnamefont {R.~M.}\ \bibnamefont {Wentzcovitch}},\
  }\href {\doibase 10.1088/0953-8984/21/39/395502} {\bibfield  {journal}
  {\bibinfo  {journal} {J. Phys.: Condens. Matter}\ }\textbf {\bibinfo {volume}
  {21}},\ \bibinfo {pages} {395502} (\bibinfo {year} {2009})}\BibitemShut
  {NoStop}%
\bibitem [{\citenamefont {Giannozzi}\ \emph {et~al.}(2017)\citenamefont
  {Giannozzi}, \citenamefont {Andreussi}, \citenamefont {Brumme}, \citenamefont
  {Bunau}, \citenamefont {Nardelli}, \citenamefont {Calandra}, \citenamefont
  {Car}, \citenamefont {Cavazzoni}, \citenamefont {{D Ceresoli}}, \citenamefont
  {Cococcioni}, \citenamefont {Colonna}, \citenamefont {Carnimeo},
  \citenamefont {Corso}, \citenamefont {de~Gironcoli}, \citenamefont {Delugas},
  \citenamefont {Jr}, \citenamefont {{A Ferretti}}, \citenamefont {Floris},
  \citenamefont {Fratesi}, \citenamefont {Fugallo}, \citenamefont {Gebauer},
  \citenamefont {Gerstmann}, \citenamefont {Giustino}, \citenamefont {Gorni},
  \citenamefont {Jia}, \citenamefont {Kawamura}, \citenamefont {{H-Y Ko}},
  \citenamefont {Kokalj}, \citenamefont {K\"u{\c c}\"ukbenli}, \citenamefont
  {Lazzeri}, \citenamefont {Marsili}, \citenamefont {Marzari}, \citenamefont
  {Mauri}, \citenamefont {Nguyen}, \citenamefont {Nguyen}, \citenamefont {{A
  Otero-de-la-Roza}}, \citenamefont {Paulatto}, \citenamefont {Ponc\'e},
  \citenamefont {Rocca}, \citenamefont {Sabatini}, \citenamefont {Santra},
  \citenamefont {Schlipf}, \citenamefont {Seitsonen}, \citenamefont {Smogunov},
  \citenamefont {{I Timrov}}, \citenamefont {Thonhauser}, \citenamefont
  {Umari}, \citenamefont {Vast}, \citenamefont {Wu},\ and\ \citenamefont
  {Baroni}}]{espresso_updated}%
  \BibitemOpen
  \bibfield  {author} {\bibinfo {author} {\bibfnamefont {P.}~\bibnamefont
  {Giannozzi}}, \bibinfo {author} {\bibfnamefont {O.}~\bibnamefont
  {Andreussi}}, \bibinfo {author} {\bibfnamefont {T.}~\bibnamefont {Brumme}},
  \bibinfo {author} {\bibfnamefont {O.}~\bibnamefont {Bunau}}, \bibinfo
  {author} {\bibfnamefont {M.~B.}\ \bibnamefont {Nardelli}}, \bibinfo {author}
  {\bibfnamefont {M.}~\bibnamefont {Calandra}}, \bibinfo {author}
  {\bibfnamefont {R.}~\bibnamefont {Car}}, \bibinfo {author} {\bibfnamefont
  {C.}~\bibnamefont {Cavazzoni}}, \bibinfo {author} {\bibnamefont {{D
  Ceresoli}}}, \bibinfo {author} {\bibfnamefont {M.}~\bibnamefont
  {Cococcioni}}, \bibinfo {author} {\bibfnamefont {N.}~\bibnamefont {Colonna}},
  \bibinfo {author} {\bibfnamefont {I.}~\bibnamefont {Carnimeo}}, \bibinfo
  {author} {\bibfnamefont {A.~D.}\ \bibnamefont {Corso}}, \bibinfo {author}
  {\bibfnamefont {S.}~\bibnamefont {de~Gironcoli}}, \bibinfo {author}
  {\bibfnamefont {P.}~\bibnamefont {Delugas}}, \bibinfo {author} {\bibfnamefont
  {R.~A.~D.}\ \bibnamefont {Jr}}, \bibinfo {author} {\bibnamefont {{A
  Ferretti}}}, \bibinfo {author} {\bibfnamefont {A.}~\bibnamefont {Floris}},
  \bibinfo {author} {\bibfnamefont {G.}~\bibnamefont {Fratesi}}, \bibinfo
  {author} {\bibfnamefont {G.}~\bibnamefont {Fugallo}}, \bibinfo {author}
  {\bibfnamefont {R.}~\bibnamefont {Gebauer}}, \bibinfo {author} {\bibfnamefont
  {U.}~\bibnamefont {Gerstmann}}, \bibinfo {author} {\bibfnamefont
  {F.}~\bibnamefont {Giustino}}, \bibinfo {author} {\bibfnamefont
  {T.}~\bibnamefont {Gorni}}, \bibinfo {author} {\bibfnamefont
  {J.}~\bibnamefont {Jia}}, \bibinfo {author} {\bibfnamefont {M.}~\bibnamefont
  {Kawamura}}, \bibinfo {author} {\bibnamefont {{H-Y Ko}}}, \bibinfo {author}
  {\bibfnamefont {A.}~\bibnamefont {Kokalj}}, \bibinfo {author} {\bibfnamefont
  {E.}~\bibnamefont {K\"u{\c c}\"ukbenli}}, \bibinfo {author} {\bibfnamefont
  {M.}~\bibnamefont {Lazzeri}}, \bibinfo {author} {\bibfnamefont
  {M.}~\bibnamefont {Marsili}}, \bibinfo {author} {\bibfnamefont
  {N.}~\bibnamefont {Marzari}}, \bibinfo {author} {\bibfnamefont
  {F.}~\bibnamefont {Mauri}}, \bibinfo {author} {\bibfnamefont {N.~L.}\
  \bibnamefont {Nguyen}}, \bibinfo {author} {\bibfnamefont {H.-V.}\
  \bibnamefont {Nguyen}}, \bibinfo {author} {\bibnamefont {{A
  Otero-de-la-Roza}}}, \bibinfo {author} {\bibfnamefont {L.}~\bibnamefont
  {Paulatto}}, \bibinfo {author} {\bibfnamefont {S.}~\bibnamefont {Ponc\'e}},
  \bibinfo {author} {\bibfnamefont {D.}~\bibnamefont {Rocca}}, \bibinfo
  {author} {\bibfnamefont {R.}~\bibnamefont {Sabatini}}, \bibinfo {author}
  {\bibfnamefont {B.}~\bibnamefont {Santra}}, \bibinfo {author} {\bibfnamefont
  {M.}~\bibnamefont {Schlipf}}, \bibinfo {author} {\bibfnamefont {A.~P.}\
  \bibnamefont {Seitsonen}}, \bibinfo {author} {\bibfnamefont {A.}~\bibnamefont
  {Smogunov}}, \bibinfo {author} {\bibnamefont {{I Timrov}}}, \bibinfo {author}
  {\bibfnamefont {T.}~\bibnamefont {Thonhauser}}, \bibinfo {author}
  {\bibfnamefont {P.}~\bibnamefont {Umari}}, \bibinfo {author} {\bibfnamefont
  {N.}~\bibnamefont {Vast}}, \bibinfo {author} {\bibfnamefont {X.}~\bibnamefont
  {Wu}}, \ and\ \bibinfo {author} {\bibfnamefont {S.}~\bibnamefont {Baroni}},\
  }\href {\doibase 10.1088/1361-648X/aa8f79} {\bibfield  {journal} {\bibinfo
  {journal} {J. Phys.: Condens. Matter}\ }\textbf {\bibinfo {volume} {29}},\
  \bibinfo {pages} {465901} (\bibinfo {year} {2017})}\BibitemShut {NoStop}%
\bibitem [{Note2()}]{Note2}%
  \BibitemOpen
  \bibinfo {note} {Because the first-principles calculation usually contains
  more than $M$ bands, we need to choose which bands should be represented by
  the tight-binding model.}\BibitemShut {Stop}%
\bibitem [{\citenamefont {Nelder}\ and\ \citenamefont
  {Mead}(1965)}]{nelderSimplexMethodFunction1965}%
  \BibitemOpen
  \bibfield  {author} {\bibinfo {author} {\bibfnamefont {J.~A.}\ \bibnamefont
  {Nelder}}\ and\ \bibinfo {author} {\bibfnamefont {R.}~\bibnamefont {Mead}},\
  }\href {\doibase 10.1093/comjnl/7.4.308} {\bibfield  {journal} {\bibinfo
  {journal} {Comput J}\ }\textbf {\bibinfo {volume} {7}},\ \bibinfo {pages}
  {308} (\bibinfo {year} {1965})}\BibitemShut {NoStop}%
\bibitem [{Note3()}]{Note3}%
  \BibitemOpen
  \bibinfo {note} {The constraint is implemented by assigning an infinite value
  of $\Delta $ to invalid energy windows.}\BibitemShut {Stop}%
\bibitem [{Note4()}]{Note4}%
  \BibitemOpen
  \bibinfo {note} {For storing the workflow in the AiiDA database, it needs to
  be converted into an AiiDA data type. We chose to convert it into a string
  containing the fully qualified class name, from which we import the workflow
  when needed.}\BibitemShut {Stop}%
\bibitem [{\citenamefont {Gresch}()}]{greschTBmodelsDocumentation}%
  \BibitemOpen
  \bibfield  {author} {\bibinfo {author} {\bibfnamefont {D.}~\bibnamefont
  {Gresch}},\ }\href@noop {} {\enquote {\bibinfo {title} {{{TBmodels}}
  documentation},}\ }\bibinfo {howpublished}
  {http://z2pack.ethz.ch/tbmodels}\BibitemShut {NoStop}%
\bibitem [{\citenamefont {Alicea}(2010)}]{aliceaMajoranaFermionsTunable2010}%
  \BibitemOpen
  \bibfield  {author} {\bibinfo {author} {\bibfnamefont {J.}~\bibnamefont
  {Alicea}},\ }\href {\doibase 10.1103/PhysRevB.81.125318} {\bibfield
  {journal} {\bibinfo  {journal} {Phys. Rev. B}\ }\textbf {\bibinfo {volume}
  {81}},\ \bibinfo {pages} {125318} (\bibinfo {year} {2010})}\BibitemShut
  {NoStop}%
\bibitem [{\citenamefont {Pikulin}\ \emph {et~al.}(2012)\citenamefont
  {Pikulin}, \citenamefont {Dahlhaus}, \citenamefont {Wimmer}, \citenamefont
  {Schomerus},\ and\ \citenamefont
  {Beenakker}}]{pikulinZerovoltageConductancePeak2012a}%
  \BibitemOpen
  \bibfield  {author} {\bibinfo {author} {\bibfnamefont {D.~I.}\ \bibnamefont
  {Pikulin}}, \bibinfo {author} {\bibfnamefont {J.~P.}\ \bibnamefont
  {Dahlhaus}}, \bibinfo {author} {\bibfnamefont {M.}~\bibnamefont {Wimmer}},
  \bibinfo {author} {\bibfnamefont {H.}~\bibnamefont {Schomerus}}, \ and\
  \bibinfo {author} {\bibfnamefont {C.~W.~J.}\ \bibnamefont {Beenakker}},\
  }\href {\doibase 10.1088/1367-2630/14/12/125011} {\bibfield  {journal}
  {\bibinfo  {journal} {New J. Phys.}\ }\textbf {\bibinfo {volume} {14}},\
  \bibinfo {pages} {125011} (\bibinfo {year} {2012})}\BibitemShut {NoStop}%
\bibitem [{\citenamefont {Alicea}(2012)}]{aliceaNewDirectionsPursuit2012}%
  \BibitemOpen
  \bibfield  {author} {\bibinfo {author} {\bibfnamefont {J.}~\bibnamefont
  {Alicea}},\ }\href {\doibase 10.1088/0034-4885/75/7/076501} {\bibfield
  {journal} {\bibinfo  {journal} {Rep. Prog. Phys.}\ }\textbf {\bibinfo
  {volume} {75}},\ \bibinfo {pages} {076501} (\bibinfo {year}
  {2012})}\BibitemShut {NoStop}%
\bibitem [{\citenamefont {Mi}\ \emph {et~al.}(2013)\citenamefont {Mi},
  \citenamefont {Pikulin}, \citenamefont {Wimmer},\ and\ \citenamefont
  {Beenakker}}]{miProposalDetectionBraiding2013}%
  \BibitemOpen
  \bibfield  {author} {\bibinfo {author} {\bibfnamefont {S.}~\bibnamefont
  {Mi}}, \bibinfo {author} {\bibfnamefont {D.~I.}\ \bibnamefont {Pikulin}},
  \bibinfo {author} {\bibfnamefont {M.}~\bibnamefont {Wimmer}}, \ and\ \bibinfo
  {author} {\bibfnamefont {C.~W.~J.}\ \bibnamefont {Beenakker}},\ }\href
  {\doibase 10.1103/PhysRevB.87.241405} {\bibfield  {journal} {\bibinfo
  {journal} {Phys. Rev. B}\ }\textbf {\bibinfo {volume} {87}},\ \bibinfo
  {pages} {241405} (\bibinfo {year} {2013})}\BibitemShut {NoStop}%
\bibitem [{\citenamefont
  {Beenakker}(2013)}]{beenakkerSearchMajoranaFermions2013}%
  \BibitemOpen
  \bibfield  {author} {\bibinfo {author} {\bibfnamefont {C.}~\bibnamefont
  {Beenakker}},\ }\href {\doibase 10.1146/annurev-conmatphys-030212-184337}
  {\bibfield  {journal} {\bibinfo  {journal} {Annu. Rev. Condens. Matter
  Phys.}\ }\textbf {\bibinfo {volume} {4}},\ \bibinfo {pages} {113} (\bibinfo
  {year} {2013})}\BibitemShut {NoStop}%
\bibitem [{\citenamefont
  {Kitaev}(2003)}]{kitaevFaulttolerantQuantumComputation2003}%
  \BibitemOpen
  \bibfield  {author} {\bibinfo {author} {\bibfnamefont {A.~Y.}\ \bibnamefont
  {Kitaev}},\ }\href {\doibase 10.1016/S0003-4916(02)00018-0} {\bibfield
  {journal} {\bibinfo  {journal} {Annals of Physics}\ }\textbf {\bibinfo
  {volume} {303}},\ \bibinfo {pages} {2} (\bibinfo {year} {2003})}\BibitemShut
  {NoStop}%
\bibitem [{\citenamefont {Deng}\ \emph {et~al.}(2016)\citenamefont {Deng},
  \citenamefont {Vaitiek\.{e}nas}, \citenamefont {Hansen}, \citenamefont
  {Danon}, \citenamefont {Leijnse}, \citenamefont {Flensberg}, \citenamefont
  {Nyg\aa{}rd}, \citenamefont {Krogstrup},\ and\ \citenamefont
  {Marcus}}]{dengMajoranaBoundState2016}%
  \BibitemOpen
  \bibfield  {author} {\bibinfo {author} {\bibfnamefont {M.~T.}\ \bibnamefont
  {Deng}}, \bibinfo {author} {\bibfnamefont {S.}~\bibnamefont
  {Vaitiek\.{e}nas}}, \bibinfo {author} {\bibfnamefont {E.~B.}\ \bibnamefont
  {Hansen}}, \bibinfo {author} {\bibfnamefont {J.}~\bibnamefont {Danon}},
  \bibinfo {author} {\bibfnamefont {M.}~\bibnamefont {Leijnse}}, \bibinfo
  {author} {\bibfnamefont {K.}~\bibnamefont {Flensberg}}, \bibinfo {author}
  {\bibfnamefont {J.}~\bibnamefont {Nyg\aa{}rd}}, \bibinfo {author}
  {\bibfnamefont {P.}~\bibnamefont {Krogstrup}}, \ and\ \bibinfo {author}
  {\bibfnamefont {C.~M.}\ \bibnamefont {Marcus}},\ }\href {\doibase
  10.1126/science.aaf3961} {\bibfield  {journal} {\bibinfo  {journal}
  {Science}\ }\textbf {\bibinfo {volume} {354}},\ \bibinfo {pages} {1557}
  (\bibinfo {year} {2016})}\BibitemShut {NoStop}%
\bibitem [{\citenamefont {Kane}\ and\ \citenamefont
  {Mele}(2005)}]{kaneQuantumSpinHall2005}%
  \BibitemOpen
  \bibfield  {author} {\bibinfo {author} {\bibfnamefont {C.~L.}\ \bibnamefont
  {Kane}}\ and\ \bibinfo {author} {\bibfnamefont {E.~J.}\ \bibnamefont
  {Mele}},\ }\href {\doibase 10.1103/PhysRevLett.95.226801} {\bibfield
  {journal} {\bibinfo  {journal} {Phys. Rev. Lett.}\ }\textbf {\bibinfo
  {volume} {95}},\ \bibinfo {pages} {226801} (\bibinfo {year}
  {2005})}\BibitemShut {NoStop}%
\bibitem [{\citenamefont {Liu}\ \emph {et~al.}(2008)\citenamefont {Liu},
  \citenamefont {Hughes}, \citenamefont {Qi}, \citenamefont {Wang},\ and\
  \citenamefont {Zhang}}]{liuQuantumSpinHall2008}%
  \BibitemOpen
  \bibfield  {author} {\bibinfo {author} {\bibfnamefont {C.}~\bibnamefont
  {Liu}}, \bibinfo {author} {\bibfnamefont {T.~L.}\ \bibnamefont {Hughes}},
  \bibinfo {author} {\bibfnamefont {X.-L.}\ \bibnamefont {Qi}}, \bibinfo
  {author} {\bibfnamefont {K.}~\bibnamefont {Wang}}, \ and\ \bibinfo {author}
  {\bibfnamefont {S.-C.}\ \bibnamefont {Zhang}},\ }\href {\doibase
  10.1103/PhysRevLett.100.236601} {\bibfield  {journal} {\bibinfo  {journal}
  {Phys. Rev. Lett.}\ }\textbf {\bibinfo {volume} {100}},\ \bibinfo {pages}
  {236601} (\bibinfo {year} {2008})}\BibitemShut {NoStop}%
\bibitem [{\citenamefont {Mourik}\ \emph {et~al.}(2012)\citenamefont {Mourik},
  \citenamefont {Zuo}, \citenamefont {Frolov}, \citenamefont {Plissard},
  \citenamefont {Bakkers},\ and\ \citenamefont
  {Kouwenhoven}}]{mourikSignaturesMajoranaFermions2012}%
  \BibitemOpen
  \bibfield  {author} {\bibinfo {author} {\bibfnamefont {V.}~\bibnamefont
  {Mourik}}, \bibinfo {author} {\bibfnamefont {K.}~\bibnamefont {Zuo}},
  \bibinfo {author} {\bibfnamefont {S.~M.}\ \bibnamefont {Frolov}}, \bibinfo
  {author} {\bibfnamefont {S.~R.}\ \bibnamefont {Plissard}}, \bibinfo {author}
  {\bibfnamefont {E.~P. a.~M.}\ \bibnamefont {Bakkers}}, \ and\ \bibinfo
  {author} {\bibfnamefont {L.~P.}\ \bibnamefont {Kouwenhoven}},\ }\href
  {\doibase 10.1126/science.1222360} {\bibfield  {journal} {\bibinfo  {journal}
  {Science}\ }\textbf {\bibinfo {volume} {336}},\ \bibinfo {pages} {1003}
  (\bibinfo {year} {2012})}\BibitemShut {NoStop}%
\bibitem [{\citenamefont {Churchill}\ \emph {et~al.}(2013)\citenamefont
  {Churchill}, \citenamefont {Fatemi}, \citenamefont {{Grove-Rasmussen}},
  \citenamefont {Deng}, \citenamefont {Caroff}, \citenamefont {Xu},\ and\
  \citenamefont
  {Marcus}}]{churchillSuperconductornanowireDevicesTunneling2013}%
  \BibitemOpen
  \bibfield  {author} {\bibinfo {author} {\bibfnamefont {H.~O.~H.}\
  \bibnamefont {Churchill}}, \bibinfo {author} {\bibfnamefont {V.}~\bibnamefont
  {Fatemi}}, \bibinfo {author} {\bibfnamefont {K.}~\bibnamefont
  {{Grove-Rasmussen}}}, \bibinfo {author} {\bibfnamefont {M.~T.}\ \bibnamefont
  {Deng}}, \bibinfo {author} {\bibfnamefont {P.}~\bibnamefont {Caroff}},
  \bibinfo {author} {\bibfnamefont {H.~Q.}\ \bibnamefont {Xu}}, \ and\ \bibinfo
  {author} {\bibfnamefont {C.~M.}\ \bibnamefont {Marcus}},\ }\href {\doibase
  10.1103/PhysRevB.87.241401} {\bibfield  {journal} {\bibinfo  {journal} {Phys.
  Rev. B}\ }\textbf {\bibinfo {volume} {87}},\ \bibinfo {pages} {241401}
  (\bibinfo {year} {2013})}\BibitemShut {NoStop}%
\bibitem [{\citenamefont {Becke}(1993)}]{hybrids}%
  \BibitemOpen
  \bibfield  {author} {\bibinfo {author} {\bibfnamefont {A.~D.}\ \bibnamefont
  {Becke}},\ }\href {\doibase 10.1063/1.464913} {\bibfield  {journal} {\bibinfo
   {journal} {The Journal of Chemical Physics}\ }\textbf {\bibinfo {volume}
  {98}},\ \bibinfo {pages} {5648} (\bibinfo {year} {1993})}\BibitemShut
  {NoStop}%
\bibitem [{\citenamefont {Hedin}(1965)}]{gw}%
  \BibitemOpen
  \bibfield  {author} {\bibinfo {author} {\bibfnamefont {L.}~\bibnamefont
  {Hedin}},\ }\href {\doibase 10.1103/PhysRev.139.A796} {\bibfield  {journal}
  {\bibinfo  {journal} {Phys. Rev.}\ }\textbf {\bibinfo {volume} {139}},\
  \bibinfo {pages} {A796} (\bibinfo {year} {1965})}\BibitemShut {NoStop}%
\bibitem [{\citenamefont {Kane}(1966)}]{kaneChapterMethod1966}%
  \BibitemOpen
  \bibfield  {author} {\bibinfo {author} {\bibfnamefont {E.~O.}\ \bibnamefont
  {Kane}},\ }in\ \href {\doibase 10.1016/S0080-8784(08)62376-5} {\emph
  {\bibinfo {booktitle} {Semiconductors and {{Semimetals}}}}},\ \bibinfo
  {series} {Semiconductors and Semimetals}, Vol.~\bibinfo {volume} {1},\
  \bibinfo {editor} {edited by\ \bibinfo {editor} {\bibfnamefont {R.~K.}\
  \bibnamefont {Willardson}}\ and\ \bibinfo {editor} {\bibfnamefont {A.~C.}\
  \bibnamefont {Beer}}}\ (\bibinfo  {publisher} {{Elsevier}},\ \bibinfo {year}
  {1966})\ pp.\ \bibinfo {pages} {75--100}\BibitemShut {NoStop}%
\bibitem [{\citenamefont {Slater}\ and\ \citenamefont
  {Koster}(1954)}]{slaterSimplifiedLCAOMethod1954}%
  \BibitemOpen
  \bibfield  {author} {\bibinfo {author} {\bibfnamefont {J.~C.}\ \bibnamefont
  {Slater}}\ and\ \bibinfo {author} {\bibfnamefont {G.~F.}\ \bibnamefont
  {Koster}},\ }\href {\doibase 10.1103/PhysRev.94.1498} {\bibfield  {journal}
  {\bibinfo  {journal} {Phys. Rev.}\ }\textbf {\bibinfo {volume} {94}},\
  \bibinfo {pages} {1498} (\bibinfo {year} {1954})}\BibitemShut {NoStop}%
\bibitem [{\citenamefont {Tan}\ \emph {et~al.}(2013)\citenamefont {Tan},
  \citenamefont {Povolotskyi}, \citenamefont {Kubis}, \citenamefont {He},
  \citenamefont {Jiang}, \citenamefont {Klimeck},\ and\ \citenamefont
  {Boykin}}]{tanEmpiricalTightBinding2013}%
  \BibitemOpen
  \bibfield  {author} {\bibinfo {author} {\bibfnamefont {Y.}~\bibnamefont
  {Tan}}, \bibinfo {author} {\bibfnamefont {M.}~\bibnamefont {Povolotskyi}},
  \bibinfo {author} {\bibfnamefont {T.}~\bibnamefont {Kubis}}, \bibinfo
  {author} {\bibfnamefont {Y.}~\bibnamefont {He}}, \bibinfo {author}
  {\bibfnamefont {Z.}~\bibnamefont {Jiang}}, \bibinfo {author} {\bibfnamefont
  {G.}~\bibnamefont {Klimeck}}, \ and\ \bibinfo {author} {\bibfnamefont
  {T.~B.}\ \bibnamefont {Boykin}},\ }\href {\doibase 10.1007/s10825-013-0436-0}
  {\bibfield  {journal} {\bibinfo  {journal} {J Comput Electron}\ }\textbf
  {\bibinfo {volume} {12}},\ \bibinfo {pages} {56} (\bibinfo {year}
  {2013})}\BibitemShut {NoStop}%
\bibitem [{\citenamefont {Tan}\ \emph {et~al.}(2015)\citenamefont {Tan},
  \citenamefont {Povolotskyi}, \citenamefont {Kubis}, \citenamefont {Boykin},\
  and\ \citenamefont {Klimeck}}]{tanTightbindingAnalysisSi2015}%
  \BibitemOpen
  \bibfield  {author} {\bibinfo {author} {\bibfnamefont {Y.~P.}\ \bibnamefont
  {Tan}}, \bibinfo {author} {\bibfnamefont {M.}~\bibnamefont {Povolotskyi}},
  \bibinfo {author} {\bibfnamefont {T.}~\bibnamefont {Kubis}}, \bibinfo
  {author} {\bibfnamefont {T.~B.}\ \bibnamefont {Boykin}}, \ and\ \bibinfo
  {author} {\bibfnamefont {G.}~\bibnamefont {Klimeck}},\ }\href {\doibase
  10.1103/PhysRevB.92.085301} {\bibfield  {journal} {\bibinfo  {journal} {Phys.
  Rev. B}\ }\textbf {\bibinfo {volume} {92}},\ \bibinfo {pages} {085301}
  (\bibinfo {year} {2015})}\BibitemShut {NoStop}%
\bibitem [{\citenamefont {Tan}\ \emph {et~al.}(2016)\citenamefont {Tan},
  \citenamefont {Povolotskyi}, \citenamefont {Kubis}, \citenamefont {Boykin},\
  and\ \citenamefont {Klimeck}}]{tanTransferableTightbindingModel2016}%
  \BibitemOpen
  \bibfield  {author} {\bibinfo {author} {\bibfnamefont {Y.}~\bibnamefont
  {Tan}}, \bibinfo {author} {\bibfnamefont {M.}~\bibnamefont {Povolotskyi}},
  \bibinfo {author} {\bibfnamefont {T.}~\bibnamefont {Kubis}}, \bibinfo
  {author} {\bibfnamefont {T.~B.}\ \bibnamefont {Boykin}}, \ and\ \bibinfo
  {author} {\bibfnamefont {G.}~\bibnamefont {Klimeck}},\ }\href {\doibase
  10.1103/PhysRevB.94.045311} {\bibfield  {journal} {\bibinfo  {journal} {Phys.
  Rev. B}\ }\textbf {\bibinfo {volume} {94}},\ \bibinfo {pages} {045311}
  (\bibinfo {year} {2016})}\BibitemShut {NoStop}%
\bibitem [{\citenamefont {Jancu}\ \emph {et~al.}(1998)\citenamefont {Jancu},
  \citenamefont {Scholz}, \citenamefont {Beltram},\ and\ \citenamefont
  {Bassani}}]{jancuEmpiricalMathrmspdsTightbinding1998}%
  \BibitemOpen
  \bibfield  {author} {\bibinfo {author} {\bibfnamefont {J.-M.}\ \bibnamefont
  {Jancu}}, \bibinfo {author} {\bibfnamefont {R.}~\bibnamefont {Scholz}},
  \bibinfo {author} {\bibfnamefont {F.}~\bibnamefont {Beltram}}, \ and\
  \bibinfo {author} {\bibfnamefont {F.}~\bibnamefont {Bassani}},\ }\href
  {\doibase 10.1103/PhysRevB.57.6493} {\bibfield  {journal} {\bibinfo
  {journal} {Phys. Rev. B}\ }\textbf {\bibinfo {volume} {57}},\ \bibinfo
  {pages} {6493} (\bibinfo {year} {1998})}\BibitemShut {NoStop}%
\bibitem [{\citenamefont {Boykin}(1997)}]{boykinImprovedFitsEffective1997}%
  \BibitemOpen
  \bibfield  {author} {\bibinfo {author} {\bibfnamefont {T.~B.}\ \bibnamefont
  {Boykin}},\ }\href {\doibase 10.1103/PhysRevB.56.9613} {\bibfield  {journal}
  {\bibinfo  {journal} {Phys. Rev. B}\ }\textbf {\bibinfo {volume} {56}},\
  \bibinfo {pages} {9613} (\bibinfo {year} {1997})}\BibitemShut {NoStop}%
\bibitem [{\citenamefont {Kresse}\ and\ \citenamefont
  {Furthm\"uller}(1996)}]{vasp}%
  \BibitemOpen
  \bibfield  {author} {\bibinfo {author} {\bibfnamefont {G.}~\bibnamefont
  {Kresse}}\ and\ \bibinfo {author} {\bibfnamefont {J.}~\bibnamefont
  {Furthm\"uller}},\ }\href {\doibase 10.1103/PhysRevB.54.11169} {\bibfield
  {journal} {\bibinfo  {journal} {Phys. Rev. B}\ }\textbf {\bibinfo {volume}
  {54}},\ \bibinfo {pages} {11169} (\bibinfo {year} {1996})}\BibitemShut
  {NoStop}%
\bibitem [{\citenamefont {Heyd}\ \emph {et~al.}(2003)\citenamefont {Heyd},
  \citenamefont {Scuseria},\ and\ \citenamefont {Ernzerhof}}]{hse}%
  \BibitemOpen
  \bibfield  {author} {\bibinfo {author} {\bibfnamefont {J.}~\bibnamefont
  {Heyd}}, \bibinfo {author} {\bibfnamefont {G.~E.}\ \bibnamefont {Scuseria}},
  \ and\ \bibinfo {author} {\bibfnamefont {M.}~\bibnamefont {Ernzerhof}},\
  }\href {\doibase 10.1063/1.1564060} {\bibfield  {journal} {\bibinfo
  {journal} {The Journal of Chemical Physics}\ }\textbf {\bibinfo {volume}
  {118}},\ \bibinfo {pages} {8207} (\bibinfo {year} {2003})}\BibitemShut
  {NoStop}%
\bibitem [{\citenamefont {Heyd}\ and\ \citenamefont
  {Scuseria}(2004)}]{heydEfficientHybridDensity2004}%
  \BibitemOpen
  \bibfield  {author} {\bibinfo {author} {\bibfnamefont {J.}~\bibnamefont
  {Heyd}}\ and\ \bibinfo {author} {\bibfnamefont {G.~E.}\ \bibnamefont
  {Scuseria}},\ }\href {\doibase 10.1063/1.1760074} {\bibfield  {journal}
  {\bibinfo  {journal} {The Journal of Chemical Physics}\ }\textbf {\bibinfo
  {volume} {121}},\ \bibinfo {pages} {1187} (\bibinfo {year}
  {2004})}\BibitemShut {NoStop}%
\bibitem [{\citenamefont {Perdew}\ \emph {et~al.}(1996)\citenamefont {Perdew},
  \citenamefont {Burke},\ and\ \citenamefont {Ernzerhof}}]{pbe}%
  \BibitemOpen
  \bibfield  {author} {\bibinfo {author} {\bibfnamefont {J.~P.}\ \bibnamefont
  {Perdew}}, \bibinfo {author} {\bibfnamefont {K.}~\bibnamefont {Burke}}, \
  and\ \bibinfo {author} {\bibfnamefont {M.}~\bibnamefont {Ernzerhof}},\ }\href
  {\doibase 10.1103/PhysRevLett.77.3865} {\bibfield  {journal} {\bibinfo
  {journal} {Phys. Rev. Lett.}\ }\textbf {\bibinfo {volume} {77}},\ \bibinfo
  {pages} {3865} (\bibinfo {year} {1996})}\BibitemShut {NoStop}%
\bibitem [{\citenamefont {Kim}\ \emph {et~al.}(2010)\citenamefont {Kim},
  \citenamefont {Marsman}, \citenamefont {Kresse}, \citenamefont {Tran},\ and\
  \citenamefont {Blaha}}]{kimEfficientBandStructure2010}%
  \BibitemOpen
  \bibfield  {author} {\bibinfo {author} {\bibfnamefont {Y.-S.}\ \bibnamefont
  {Kim}}, \bibinfo {author} {\bibfnamefont {M.}~\bibnamefont {Marsman}},
  \bibinfo {author} {\bibfnamefont {G.}~\bibnamefont {Kresse}}, \bibinfo
  {author} {\bibfnamefont {F.}~\bibnamefont {Tran}}, \ and\ \bibinfo {author}
  {\bibfnamefont {P.}~\bibnamefont {Blaha}},\ }\href {\doibase
  10.1103/PhysRevB.82.205212} {\bibfield  {journal} {\bibinfo  {journal} {Phys.
  Rev. B}\ }\textbf {\bibinfo {volume} {82}},\ \bibinfo {pages} {205212}
  (\bibinfo {year} {2010})}\BibitemShut {NoStop}%
\bibitem [{\citenamefont {Madelung}\ \emph {et~al.}(2002)\citenamefont
  {Madelung}, \citenamefont {R\"ossler},\ and\ \citenamefont
  {Schulz}}]{madelungCondensedMatterGroup2002}%
  \BibitemOpen
  \bibfield  {author} {\bibinfo {author} {\bibfnamefont {O.}~\bibnamefont
  {Madelung}}, \bibinfo {author} {\bibfnamefont {U.}~\bibnamefont {R\"ossler}},
  \ and\ \bibinfo {author} {\bibfnamefont {M.}~\bibnamefont {Schulz}},\
  }\href@noop {} {\  (\bibinfo {year} {2002})}\BibitemShut {NoStop}%
\bibitem [{\citenamefont {H\"auselmann}()}]{hauselmannAiiDAVASPDocumentation}%
  \BibitemOpen
  \bibfield  {author} {\bibinfo {author} {\bibfnamefont {R.}~\bibnamefont
  {H\"auselmann}},\ }\href@noop {} {\enquote {\bibinfo {title}
  {{{AiiDA}}-{{VASP}} documentation},}\ }\bibinfo {howpublished}
  {https://aiida-vasp.readthedocs.io/}\BibitemShut {NoStop}%
\bibitem [{\citenamefont {Vurgaftman}\ \emph {et~al.}(2001)\citenamefont
  {Vurgaftman}, \citenamefont {Meyer},\ and\ \citenamefont
  {{Ram-Mohan}}}]{vurgaftmanBandParametersIII2001}%
  \BibitemOpen
  \bibfield  {author} {\bibinfo {author} {\bibfnamefont {I.}~\bibnamefont
  {Vurgaftman}}, \bibinfo {author} {\bibfnamefont {J.~R.}\ \bibnamefont
  {Meyer}}, \ and\ \bibinfo {author} {\bibfnamefont {L.~R.}\ \bibnamefont
  {{Ram-Mohan}}},\ }\href {\doibase 10.1063/1.1368156} {\bibfield  {journal}
  {\bibinfo  {journal} {Journal of Applied Physics}\ }\textbf {\bibinfo
  {volume} {89}},\ \bibinfo {pages} {5815} (\bibinfo {year}
  {2001})}\BibitemShut {NoStop}%
\bibitem [{\citenamefont {Graf}\ and\ \citenamefont
  {Vogl}(1995)}]{grafElectromagneticFieldsDielectric1995}%
  \BibitemOpen
  \bibfield  {author} {\bibinfo {author} {\bibfnamefont {M.}~\bibnamefont
  {Graf}}\ and\ \bibinfo {author} {\bibfnamefont {P.}~\bibnamefont {Vogl}},\
  }\href {\doibase 10.1103/PhysRevB.51.4940} {\bibfield  {journal} {\bibinfo
  {journal} {Phys. Rev. B}\ }\textbf {\bibinfo {volume} {51}},\ \bibinfo
  {pages} {4940} (\bibinfo {year} {1995})}\BibitemShut {NoStop}%
\bibitem [{Note5()}]{Note5}%
  \BibitemOpen
  \bibinfo {note} {See Supplemental Material at [URL] for tables containing the
  band energies at high-symmetry points for different values of
  strain.}\BibitemShut {Stop}%
\bibitem [{Note6()}]{Note6}%
  \BibitemOpen
  \bibinfo {note} {See Supplemental Material at [URL] for an full export of the
  AiiDA database containing the calculations of strained InSb, InAs, and GaSb
  tight-binding models.}\BibitemShut {Stop}%
\bibitem [{Note7()}]{Note7}%
  \BibitemOpen
  \bibinfo {note} {See Supplemental Material at [URL] for an archive containing
  the $195$ strained tight-binding models of InSb, InAs, and GaSb.}\BibitemShut
  {Stop}%
\bibitem [{\citenamefont {Yusufaly}\ \emph {et~al.}(2013)\citenamefont
  {Yusufaly}, \citenamefont {Vanderbilt},\ and\ \citenamefont
  {Coh}}]{pythtb_formalism}%
  \BibitemOpen
  \bibfield  {author} {\bibinfo {author} {\bibfnamefont {T.}~\bibnamefont
  {Yusufaly}}, \bibinfo {author} {\bibfnamefont {D.}~\bibnamefont
  {Vanderbilt}}, \ and\ \bibinfo {author} {\bibfnamefont {S.}~\bibnamefont
  {Coh}},\ }\href@noop {} {\enquote {\bibinfo {title} {Tight-{{Binding
  Formalism}} in the {{Context}} of the {{PythTB Package}}},}\ }\bibinfo
  {howpublished}
  {http://www.physics.rutgers.edu/pythtb/\_downloads/pythtb-formalism.pdf}
  (\bibinfo {year} {2013})\BibitemShut {NoStop}%
\bibitem [{\citenamefont {Litvin}\ and\ \citenamefont
  {Kopsk\'y}(2011)}]{litvinSeitzNotationSymmetry2011}%
  \BibitemOpen
  \bibfield  {author} {\bibinfo {author} {\bibfnamefont {D.~B.}\ \bibnamefont
  {Litvin}}\ and\ \bibinfo {author} {\bibfnamefont {V.}~\bibnamefont
  {Kopsk\'y}},\ }\href {\doibase 10.1107/S010876731101378X} {\bibfield
  {journal} {\bibinfo  {journal} {Acta Cryst A, Acta Cryst Sect A, Acta
  Crystallogr A, Acta Crystallogr Sect A, Acta Crystallogr A Found Crystallogr,
  Acta Crystallogr Sect A Found Crystallogr}\ }\textbf {\bibinfo {volume}
  {67}},\ \bibinfo {pages} {415} (\bibinfo {year} {2011})}\BibitemShut
  {NoStop}%
\bibitem [{Note8()}]{Note8}%
  \BibitemOpen
  \bibinfo {note} {Following the same reasoning, we can see that an arbitrary
  phase of the matrix $U_g$ does not change the result.}\BibitemShut {Stop}%
\bibitem [{\citenamefont {Sun}\ \emph {et~al.}(2010)\citenamefont {Sun},
  \citenamefont {Nishida},\ and\ \citenamefont
  {Thompson}}]{sunStrainEffectSemiconductors2010}%
  \BibitemOpen
  \bibfield  {author} {\bibinfo {author} {\bibfnamefont {Y.}~\bibnamefont
  {Sun}}, \bibinfo {author} {\bibfnamefont {T.}~\bibnamefont {Nishida}}, \ and\
  \bibinfo {author} {\bibfnamefont {S.~E.}\ \bibnamefont {Thompson}},\ }\href
  {\doibase 10.1007/978-1-4419-0552-9_4} {{\selectlanguage {english}\emph
  {\bibinfo {title} {Strain {{Effect}} in {{Semiconductors}}}}}}\ (\bibinfo
  {publisher} {{Springer, Boston, MA}},\ \bibinfo {year} {2010})\BibitemShut
  {NoStop}%
\bibitem [{\citenamefont {Ma}\ \emph {et~al.}(1993)\citenamefont {Ma},
  \citenamefont {Wang},\ and\ \citenamefont
  {Schulman}}]{maBandStructureSymmetry1993}%
  \BibitemOpen
  \bibfield  {author} {\bibinfo {author} {\bibfnamefont {Q.~M.}\ \bibnamefont
  {Ma}}, \bibinfo {author} {\bibfnamefont {K.~L.}\ \bibnamefont {Wang}}, \ and\
  \bibinfo {author} {\bibfnamefont {J.~N.}\ \bibnamefont {Schulman}},\ }\href
  {\doibase 10.1103/PhysRevB.47.1936} {\bibfield  {journal} {\bibinfo
  {journal} {Phys. Rev. B}\ }\textbf {\bibinfo {volume} {47}},\ \bibinfo
  {pages} {1936} (\bibinfo {year} {1993})}\BibitemShut {NoStop}%
\bibitem [{\citenamefont {{Van de Walle}}\ and\ \citenamefont
  {Martin}(1986)}]{vandewalleTheoreticalCalculationsHeterojunction1986}%
  \BibitemOpen
  \bibfield  {author} {\bibinfo {author} {\bibfnamefont {C.~G.}\ \bibnamefont
  {{Van de Walle}}}\ and\ \bibinfo {author} {\bibfnamefont {R.~M.}\
  \bibnamefont {Martin}},\ }\href {\doibase 10.1103/PhysRevB.34.5621}
  {\bibfield  {journal} {\bibinfo  {journal} {Phys. Rev. B}\ }\textbf {\bibinfo
  {volume} {34}},\ \bibinfo {pages} {5621} (\bibinfo {year}
  {1986})}\BibitemShut {NoStop}%
\bibitem [{\citenamefont {{S\'anchez-Dehesa}}\ \emph
  {et~al.}(1982)\citenamefont {{S\'anchez-Dehesa}}, \citenamefont {Tejedor},\
  and\ \citenamefont
  {Verg\'es}}]{sanchez-dehesaSelfconsistentCalculationInternal1982}%
  \BibitemOpen
  \bibfield  {author} {\bibinfo {author} {\bibfnamefont {J.}~\bibnamefont
  {{S\'anchez-Dehesa}}}, \bibinfo {author} {\bibfnamefont {C.}~\bibnamefont
  {Tejedor}}, \ and\ \bibinfo {author} {\bibfnamefont {J.~A.}\ \bibnamefont
  {Verg\'es}},\ }\href {\doibase 10.1103/PhysRevB.26.5960} {\bibfield
  {journal} {\bibinfo  {journal} {Phys. Rev. B}\ }\textbf {\bibinfo {volume}
  {26}},\ \bibinfo {pages} {5960} (\bibinfo {year} {1982})}\BibitemShut
  {NoStop}%
\bibitem [{\citenamefont {Madelung}(1991)}]{madelungSemiconductors1991}%
  \BibitemOpen
  \bibfield  {author} {\bibinfo {author} {\bibfnamefont {O.}~\bibnamefont
  {Madelung}},\ }\href {\doibase 10.1007/978-3-642-45681-7_2} {{\selectlanguage
  {english}\emph {\bibinfo {title} {Semiconductors}}}},\ Data in Science and
  Technology\ (\bibinfo  {publisher} {{Springer, Berlin, Heidelberg}},\
  \bibinfo {year} {1991})\BibitemShut {NoStop}%
\bibitem [{\citenamefont {Brey}\ \emph {et~al.}(1987)\citenamefont {Brey},
  \citenamefont {Christensen},\ and\ \citenamefont
  {Cardona}}]{breyDeformationPotentialsValenceband1987}%
  \BibitemOpen
  \bibfield  {author} {\bibinfo {author} {\bibfnamefont {L.}~\bibnamefont
  {Brey}}, \bibinfo {author} {\bibfnamefont {N.~E.}\ \bibnamefont
  {Christensen}}, \ and\ \bibinfo {author} {\bibfnamefont {M.}~\bibnamefont
  {Cardona}},\ }\href {\doibase 10.1103/PhysRevB.36.2638} {\bibfield  {journal}
  {\bibinfo  {journal} {Phys. Rev. B}\ }\textbf {\bibinfo {volume} {36}},\
  \bibinfo {pages} {2638} (\bibinfo {year} {1987})}\BibitemShut {NoStop}%
\bibitem [{\citenamefont {Nielsen}\ and\ \citenamefont
  {Martin}(1985)}]{nielsenStressesSemiconductorsInitio1985}%
  \BibitemOpen
  \bibfield  {author} {\bibinfo {author} {\bibfnamefont {O.~H.}\ \bibnamefont
  {Nielsen}}\ and\ \bibinfo {author} {\bibfnamefont {R.~M.}\ \bibnamefont
  {Martin}},\ }\href {\doibase 10.1103/PhysRevB.32.3792} {\bibfield  {journal}
  {\bibinfo  {journal} {Phys. Rev. B}\ }\textbf {\bibinfo {volume} {32}},\
  \bibinfo {pages} {3792} (\bibinfo {year} {1985})}\BibitemShut {NoStop}%
\end{thebibliography}%

\onecolumngrid

\appendix
\crefalias{section}{appsec}

\section{Properties of the symmetrized Hamiltonian}\label{appendix:symmetrized_tb}

The symmetrized Hamiltonian is defined as (\cref{eqn:symmetrized_hamiltonian})
\begin{equation}
\tilde{\mathcal{H}}(\vec{k}) = \frac{1}{|G|} \sum_{g \in G} D^\kvec(g) \mathcal{H}(g^{-1}\vec{k}) D^\kvec(g^{-1}).
\end{equation}

We first show that this Hamiltonian respects the symmetries in $G$. Let $g' \in G$:
\begin{gather}
D^\kvec(g')\tilde{\mathcal{H}}([g']^{-1}\vec{k}) D^\kvec([g']^{-1})
= \frac{1}{|G|} \sum\limits_{g \in G} D^\kvec(g')D^\kvec(g) \mathcal{H}(g^{-1}[g']^{-1}\vec{k}) D^\kvec(g^{-1})D^\kvec([g']^{-1}) \\\nonumber
= \frac{1}{|G|} \sum\limits_{g \in G} D^\kvec(g'g) \mathcal{H}([g'g]^{-1}\vec{k}) D^\kvec([g'g]^{-1})\underset{g''=g'g}{=}\frac{1}{|G|} \sum\limits_{g'' \in G} D^\kvec(g'') \mathcal{H}([g'']^{-1}\vec{k}) D^\kvec([g'']^{-1})
= \tilde{\mathcal{H}}(\vec{k})
\end{gather}

Also, it is easily shown that symmetrizing a Hamiltonian $\mathcal{H}_\vec{k}^\text{symm.}$ which already respects the symmetry does not change it:
\begin{equation}
\tilde{\mathcal{H}}^\text{symm.}(\vec{k}) = \frac{1}{|G|} \sum_{g \in G} D^\kvec(g) \mathcal{H}^\text{symm.}(g^{-1}\vec{k}) D^\kvec(g^{-1})
\underset{\text{\cref{eqn:symmetry_constraint}}}{=}  \frac{1}{|G|} \sum_{g \in G}  \mathcal{H}^\text{symm.}(\vec{k}) = \mathcal{H}^\text{symm.}(\vec{k}).
\end{equation}

\section{Symmetrized Hamiltonian in terms of real-space matrices}\label{app:symmetrized_H_in_real_space}

In this appendix, we show how the symmetrized Hamiltonian $\tilde{\mathcal{H}}(\kvec)$ can be expressed in terms of real-space hopping matrices $H[\Rvec]$. 
In the following, we will assume that the representation matrix $D^\kvec(g)$ of unitary operations is given in the form 
\begin{equation}\label{eqn:unitary_op}
D^\kvec(g) = e^{i \boldsymbol{\alpha}_g .\kvec} D(g) = e^{i \boldsymbol{\alpha}_g .\kvec} U_g,
\end{equation}
where $\boldsymbol{\alpha}_g$ is the translation vector and $U_g$ is a unitary matrix. For anti-unitary operations, we assume they are of the form
\begin{equation}\label{eqn:antiunitary_op}
D^\kvec(g) = e^{i \boldsymbol{\alpha}_g .\kvec} D(g) = e^{i \boldsymbol{\alpha}_g .\kvec} U_g \hat{K},
\end{equation}
where $\hat{K}$ represents complex conjugation.

First, we notice that $D_{il}^\kvec(g) \neq 0$ only if $g \tvec_l - \tvec_i \in \mathbb{Z}^d$, meaning that orbitals centered at $\tvec_l$ are mapped onto $\tvec_i$, up to a possible lattice translation. Using \cref{eqn:tight_binding,eqn:symmetrized_hamiltonian},
we can write the symmetrized Hamiltonian as
\begin{equation}\label{eqn:symmetric_h_with_r}
\tilde{\mathcal{H}}^{ij}(\kvec) = \frac{1}{|G|} \sum_{\substack{g \in G\\{l,m}\\\Rvec}} D_{il}^\kvec(g) H^{lm}[\Rvec]e^{i (g^{-1}\kvec).(\Rvec + \tvec_m - \tvec_l)} D_{mj}^\kvec(g^{-1}),
\end{equation}
where the indices $l, m$ only go over non-zero $D_{il}^\kvec(g)$ and $D_{mj}^\kvec(g^{-1})$. Writing the real-space operator for $g$ in Seitz notation~\cite{litvinSeitzNotationSymmetry2011}
\begin{equation}
g_\vec{r} = \left\{S_g \middle| \boldsymbol{\alpha}_g \right\},\qquad g_\vec{r}^{-1} = \left\{S_g^{-1} \middle| -S_g^{-1} \boldsymbol{\alpha}_g \right\},
\end{equation}
where $S_g$ is the rotational part, and $\boldsymbol{\alpha}_g$ is the translation vector of the symmetry, this means that 
\begin{gather}\label{eqn:tmp_first}
g \tvec_l - \tvec_i = S_g \tvec_l + \boldsymbol{\alpha}_g - \tvec_i \in \mathbb{Z}^d\quad
\\\label{eqn:tmp_second}
g^{-1}\tvec_j - \tvec_m = S_g^{-1} \tvec_j -  S_g^{-1} \boldsymbol{\alpha}_g - \tvec_m  \in \mathbb{Z}^d \Rightarrow \tvec_j - \boldsymbol{\alpha}_g - S_g \tvec_m \in \mathbb{Z}^d\\
\underset{-(\ref{eqn:tmp_first}) - (\ref{eqn:tmp_second})}{\Rightarrow} \vec{T}_{ij}^{ml} = S_g(\tvec_m - \tvec_l) - (\tvec_j - \tvec_i) \in \mathbb{Z}^d.
\end{gather}
Next, we must consider how $g$ acts on the reciprocal-space vector $\kvec$. For an (anti-) unitary operator, we know that 
\begin{equation}
\langle \kvec , \vec{r} \rangle = \pm \langle g \kvec, g \vec{r} \rangle,
\end{equation}
where the positive (negative) sign corresponds to the unitary (anti-unitary) case. Since $g$ acts on $\vec{r}$ with $S_g$, it follows that
\begin{gather}
\langle \kvec , \vec{r} \rangle = \pm \langle A \kvec, S_g \vec{r} \rangle \\\nonumber
\kvec^T  \vec{r} = \pm \kvec^T A^T S_g \vec{r},
\end{gather}
where $A$ is the operation which acts upon $\kvec$ when $g$ is applied. Since this is true for all $\kvec$ and $\vec{r}$, 
\begin{equation}
A = \pm (S_g^T)^{-1},
\end{equation}
and thus
\begin{equation}\label{eqn:g_inv_on_k}
g^{-1} \kvec = \pm S_g^T \kvec.
\end{equation}
For the next step, we treat the unitary and anti-unitary cases separately for clarity.

\begin{enumerate}
\item \textbf{Unitary case}

By applying \cref{eqn:g_inv_on_k} to \cref{eqn:symmetric_h_with_r}, we get
\begin{align}
\tilde{\mathcal{H}}^{ij}(\kvec)
=~& \frac{1}{|G|} \sum_{\substack{g \in G\\{l,m}\\\Rvec}} D_{il}^\kvec(g) H^{lm}[\Rvec]e^{i \left(S_g^T\kvec\right).(\Rvec + \tvec_m - \tvec_l)}D_{mj}^\kvec(g^{-1}) \\\nonumber
=~& \frac{1}{|G|} \sum_{\substack{g \in G\\{l,m}\\\Rvec}} D_{il}^\kvec(g) H^{lm}[\Rvec]e^{i \kvec.\left[ S_g(\Rvec + \tvec_m - \tvec_l)\right]}D_{mj}^\kvec(g^{-1}).
\end{align}
Applying \cref{eqn:unitary_op}, we obtain
\begin{align}
\tilde{\mathcal{H}}^{ij}(\kvec) = ~& \frac{1}{|G|} \sum_{\substack{g \in G\\{l,m}\\\Rvec}} e^{i \boldsymbol{\alpha}_g . \kvec}\left(U_g\right)_{il} H^{lm}[\Rvec]e^{i \kvec.\left[ S_g(\Rvec + \tvec_m - \tvec_l)\right]}\left(U_g^\dagger\right)_{mj} e^{-i \boldsymbol{\alpha}_g . \kvec} \\\nonumber
= ~& \frac{1}{|G|} \sum_{\substack{g \in G\\{l,m}\\\Rvec}} \left(U_g\right)_{il} H^{lm}[\Rvec]\left(U_g^\dagger\right)_{mj} e^{i \kvec.\left[ S_g(\Rvec + \tvec_m - \tvec_l)\right]} \\\nonumber
= ~& \frac{1}{|G|} \sum_{\substack{g \in G\\{l,m}\\\Rvec}} D_{il}(g) H^{lm}[\Rvec]D_{mj}(g^{-1}) e^{i \kvec.\left[ S_g(\Rvec + \tvec_m - \tvec_l)\right]}. \\\nonumber
\end{align}

\item \textbf{Anti-unitary case}

In the anti-unitary case, we get
\begin{align}\label{eqn:tmp_antiuntary}
\tilde{\mathcal{H}}^{ij}(\kvec)
=~& \frac{1}{|G|} \sum_{\substack{g \in G\\{l,m}\\\Rvec}} D_{il}^\kvec(g) H^{lm}[\Rvec]e^{i \left(-S_g^T\kvec\right).(\Rvec + \tvec_m - \tvec_l)}D_{mj}^\kvec(g^{-1}) \\\nonumber
=~& \frac{1}{|G|} \sum_{\substack{g \in G\\{l,m}\\\Rvec}} D_{il}^\kvec(g) H^{lm}[\Rvec]e^{-i \kvec.\left[ S_g(\Rvec + \tvec_m - \tvec_l)\right]}D_{mj}^\kvec(g^{-1}).
\end{align}
When applying \cref{eqn:antiunitary_op}, it is important to note that the representation of the inverse is given by
\begin{gather}\label{eqn:inverse_antiunitary_repr}
D^\kvec(g^{-1}) = \left(D^\kvec(g)\right)^{-1} = \left(e^{i \boldsymbol{\alpha}_g .\kvec} U_g \hat{K}\right)^{-1} = \hat{K} U_g^\dagger e^{-i \boldsymbol{\alpha}_g .\kvec}\\\nonumber
= e^{i \boldsymbol{\alpha}_g .\kvec} \hat{K} U_g^\dagger = e^{i \boldsymbol{\alpha}_g .\kvec} D(g^{-1}).
\end{gather}
Applying \cref{eqn:antiunitary_op,eqn:inverse_antiunitary_repr} to \cref{eqn:tmp_antiuntary}, we get
\begin{align}
\tilde{\mathcal{H}}^{ij}(\kvec) = ~& \frac{1}{|G|} \sum_{\substack{g \in G\\{l,m}\\\Rvec}} e^{i \boldsymbol{\alpha}_g . \kvec}\left(U_g\right)_{il} \hat{K} H^{lm}[\Rvec]e^{i \kvec.\left[ -S_g(\Rvec + \tvec_m - \tvec_l)\right]} e^{i \boldsymbol{\alpha}_g . \kvec}\hat{K}\left(U_g^\dagger\right)_{mj}  \\\nonumber
= ~& \frac{1}{|G|} \sum_{\substack{g \in G\\{l,m}\\\Rvec}} \left(U_g\right)_{il} \hat{K} H^{lm}[\Rvec]\hat{K} \left(U_g^\dagger\right)_{mj} e^{i \kvec.\left[ S_g(\Rvec + \tvec_m - \tvec_l)\right]} \\\nonumber
= ~& \frac{1}{|G|} \sum_{\substack{g \in G\\{l,m}\\\Rvec}} D_{il}(g) H^{lm}[\Rvec]D_{mj}(g^{-1}) e^{i \kvec.\left[ S_g(\Rvec + \tvec_m - \tvec_l)\right]}. \\\nonumber
\end{align}

\end{enumerate}

We observe that the result is the same for the unitary and anti-unitary cases, and treat them together in the following. Note that the $\kvec$ - dependent part of the representation cancels with its inverse in both cases \footnote{Following the same reasoning, we can see that an arbitrary phase of the matrix $U_g$ does not change the result.}.

Next, we substitute $\tvec_m - \tvec_l$ using $\vec{T}_{ij}^{ml}$ defined above, and define $\vec{R'} = S_g\Rvec + \vec{T}_{ij}^{ml}$. Since $\Rvec'$ is again a lattice vector, we can change the summation from $\Rvec$ to $\Rvec'$:
\begin{align}
\tilde{\mathcal{H}}^{ij}(\kvec)
= &~\frac{1}{|G|} \sum_{\substack{g \in G\\{l,m}\\\Rvec}} D_{il}(g) H^{lm}[\Rvec]D_{mj}(g^{-1})e^{i \kvec.[S_g\Rvec + \vec{T}_{ij}^{ml} + \tvec_j - \tvec_i]} \\
= &~\frac{1}{|G|} \sum_{\substack{g \in G\\{l,m}\\\Rvec'}} D_{il}(g) H^{lm}[S_g^{-1}(\Rvec' - \vec{T}_{ij}^{ml})]D_{mj}(g^{-1})e^{i \kvec.(\Rvec' + \tvec_j - \tvec_i)}.
\end{align}
Finally, we again use \cref{eqn:tight_binding} to obtain the symmetrized real-space hopping matrices
\begin{equation}
\tilde{H}^{ij}[\Rvec'] = \frac{1}{|G|} \sum_{\substack{g \in G \\{l,m}}} D_{il}(g) H^{lm}[S_g^{-1}(\Rvec' - \vec{T}_{ij}^{ml})] D_{mj}(g^{-1}).
\end{equation}

\section{AiiDA \emph{expose} functionality}\label{app:expose}

In this appendix, we illustrate how the AiiDA \emph{expose} functionality simplifies writing modular workflows. It allows implicitly forwarding input and output values of a sub-workflow instead of having to explicitly specify each value. \Cref{lst:expose_subwf} shows a simple workflow with two inputs \texttt{a} and \texttt{b}, and one output \texttt{c}. A parent workflow that only wraps this workflow is shown in \cref{lst:expose_parent_old,lst:expose_parent_new} with and without using the \emph{expose} functionality, respectively. Import statements are omitted in all listings for brevity. 

Besides reducing the boilerplate code in the parent workflow, this enables adhering to the single responsibility principle: The parent workflow does not need to change if the inputs or outputs of the wrapped workflow change, unless it directly impacts the parent workflow logic.

\begin{lstlisting}[language=Python, caption={A simple workflow with inputs \texttt{a} and \texttt{b}, and output \texttt{c}. The steps executing the workflow are omitted.}, label=lst:expose_subwf]
class SubWF(WorkChain):
    @classmethod
    def define(cls, spec):
        spec.input('a', valid_type=Int)
        spec.input('b', valid_type=Int)
        spec.output('c', valid_type=Int)
    ...
\end{lstlisting}

\begin{lstlisting}[language=Python, caption={A workflow that wraps \texttt{SubWF} by using the \emph{expose} functionality.}, label=lst:expose_parent_new]
class ParentWF(WorkChain):
    @classmethod
    def define(cls, spec):
        spec.expose_inputs(SubWF)
        spec.expose_outputs(SubWF)

        spec.outline(
            cls.invoke_subwf,
            cls.write_outputs
        )

    def invoke_subwf(self):
        return ToContext(
            sub_wf=self.submit(SubWF, **self.exposed_inputs(SubWF))
        )

    def write_outputs(self):
        self.out_many(self.exposed_outputs(self.ctx.sub_wf))
\end{lstlisting}

\begin{lstlisting}[language=Python, caption={A workflow that wraps \texttt{SubWF} without using the \emph{expose} functionality.}, label=lst:expose_parent_old]
class ParentWF(WorkChain):
    @classmethod
    def define(cls, spec):
        spec.input('a', valid_type=Int)
        spec.input('b', valid_type=Int)
        spec.output('c', valid_type=Int)

        spec.outline(
            cls.invoke_subwf,
            cls.write_outputs
        )

    def invoke_subwf(self):
        return ToContext(
            sub_wf=self.submit(SubWF, a=self.inputs.a, b=self.inputs.b)
        )

    def write_outputs(self):
        self.out('c', self.ctx.sub_wf.out.c)
\end{lstlisting}

\section{Strain tensor and strained atom position}\label{app:strain_details}
In this section, we list all the strain tensors that we used in the above simulations. Under a small homogeneous and elastic strain, the lattice vectors $\mathbf{R}$ transform (in Cartesian coordinates) into~\cite{sunStrainEffectSemiconductors2010,maBandStructureSymmetry1993}
\begin{equation}
\mathbf{R'} = (1 + \boldsymbol{\epsilon})\mathbf{R},
\end{equation}
where
\begin{eqnarray}
\boldsymbol{\epsilon}=
\left(\begin{array}{ccc}\epsilon_{xx} & \epsilon_{xy} & \epsilon_{xz} \\
\epsilon_{yx} & \epsilon_{yy} & \epsilon_{yz} \\
\epsilon_{zx} & \epsilon_{zy} & \epsilon_{zz}\end{array}\right)
\end{eqnarray}
is the strain tensor. Due to the stress-strain relation, the strain tensor under different kinds of strain can be obtained as listed below~\cite{vandewalleTheoreticalCalculationsHeterojunction1986,sanchez-dehesaSelfconsistentCalculationInternal1982,maBandStructureSymmetry1993}:

\begin{enumerate}[label=\textbf{\arabic*}.]
\item \textbf{(001) plane biaxial strain}
\begin{eqnarray}
\boldsymbol{\epsilon}^{bi}_{001}=
\left(
\begin{array}{ccc}\epsilon_{xx} &0 &0\\
0 & \epsilon_{yy} & 0 \\
0 & 0& \epsilon_{zz}\end{array}
\right)
\end{eqnarray}
where $\epsilon_{xx}=\epsilon_{yy}=\epsilon'$, $\epsilon_{zz}=-2\frac{C_{12}}{C_{11}}\epsilon'$.

\item \textbf{(110) plane biaxial strain}
\begin{eqnarray}
\boldsymbol{\epsilon}^{bi}_{110}=
\left(
\begin{array}{ccc}\epsilon_{xx} &\epsilon_{xy} &0\\
\epsilon_{xy} & \epsilon_{xx} & 0 \\
0 & 0& \epsilon_{zz}\end{array}
\right)
\end{eqnarray}
where
\begin{eqnarray}
\epsilon_{zz}&=&\epsilon'\\\nonumber
\epsilon_{xx}&=&\frac{2C_{44}-C_{12}}{2C_{44}+C_{11}+C_{12}}\epsilon'\\\nonumber
\epsilon_{xy}&=&\frac{-C_{11}-2C_{12}}{2C_{44}+C_{11}+C_{12}}\epsilon'
\end{eqnarray}

\item \textbf{(111) plane biaxial strain}
\begin{eqnarray}
\boldsymbol{\epsilon}^{bi}_{111}=
\left(
\begin{array}{ccc}\epsilon_{xx} &\epsilon_{xy} &\epsilon_{xy}\\
\epsilon_{xy} & \epsilon_{xx} & \epsilon_{xy} \\
\epsilon_{xy} & \epsilon_{xy}& \epsilon_{xx}\end{array}
\right)
\end{eqnarray}
where
\begin{eqnarray}
\epsilon_{xx}&=&\frac{4C_{44}}{4C_{44}+C_{11}+2C_{12}}\epsilon'\\\nonumber
\epsilon_{xy}&=&\frac{-C_{11}-2C_{12}}{4C_{44}+C_{11}+2C_{12}}\epsilon'
\end{eqnarray}

\item \textbf{[110] uniaxial strain}
\begin{eqnarray}
\boldsymbol{\epsilon}^{uni}_{110}=
\left(
\begin{array}{ccc}\epsilon_{xx} &\epsilon_{xy} &0\\
\epsilon_{xy} & \epsilon_{xx} & 0 \\
0 & 0& \epsilon_{zz}\end{array}
\right)
\end{eqnarray}
where
\begin{eqnarray}
\epsilon_{zz}&=&\epsilon'\\\nonumber
\epsilon_{xx}&=&-\frac{C_{11}}{2C_{12}}\epsilon'\\\nonumber
\epsilon_{xy}&=&-\frac{(C_{11}-C_{12})(C_{11}+2C_{12})}{4C_{44}C_{12}}\epsilon'
\end{eqnarray}
\end{enumerate}

In the distorted system, the position of the atoms also changes with the strain tensor. In the unstrained InAs, GaSb and InSb system, the cation is located in the (0, 0, 0) site and the anion is located at $\hat{\tau}=$(1/4, 1/4, 1/4) in primitive lattice vectors. In Cartesian coordinates, $\tau$ changes by the following~\cite{vandewalleTheoreticalCalculationsHeterojunction1986,sanchez-dehesaSelfconsistentCalculationInternal1982,maBandStructureSymmetry1993}:

\begin{enumerate}[label=\textbf{\arabic*}.]
\item \textbf{(001) biaxial strain}
\begin{eqnarray}
\hat{\tau}'= (1+\boldsymbol{\epsilon}) \hat{\tau}
\end{eqnarray}

\item \textbf{(111) biaxial strain}
\begin{equation}
\hat{\tau}'= (1+\boldsymbol{\epsilon}) \hat{\tau}- \frac{a_0}{2}\epsilon_{xy} \zeta \left(
\begin{array}{c}
1\\
1\\
1
\end{array}
\right)
 = (1+\boldsymbol{\epsilon} - 2 \epsilon_{xy} \zeta) \hat{\tau}
,
\end{equation}
where $a_0$ is the lattice constant without strain.

\item \textbf{(110) biaxial strain and [110] uniaxial strain}
\begin{equation}
\hat{\tau}'= (1+\boldsymbol{\epsilon}) \hat{\tau}- \frac{a_0}{2}\epsilon_{xy} \zeta \left(
\begin{array}{c}
0\\
0\\
1
\end{array}
\right)
 = \left(1+\boldsymbol{\epsilon} - 2 \epsilon_{xy} \zeta
 \left(\begin{array}{ccc}
 0 & 0 & 0\\
 0 & 0 & 0\\
 0 & 0 & 1
 \end{array}\right)
 \right) \hat{\tau}
.
\end{equation}
\end{enumerate}

The internal displacement  $\zeta$ and the stiffness constants $C_{11}$, $C_{12}$, $C_{44}$ of InAs, GaSb and InSb we used in the paper are listed in \cref{tab:stiffness}
\begin{table}
\begin{tabular}{ cccccc}
    \hline
    Quantity &Symbol & unit   &  InAs   & GaSb   & InSb  \\\hline
       &   $C_{11}$  &   $10^{11}  \text{dyn cm}^{-2}$        & 8.329 &8.834&6.918\\
     Elastic constant$^a$  &   $C_{12}$  &  $10^{11}  \text{dyn cm}^{-2} $        & 4.526 &4.023&3.788\\
       &   $C_{44}$  &  $10^{11}  \text{dyn cm}^{-2} $          & 3.959 &4.322&3.132\\
       \hline
       internal dis.$^b$&   $\zeta$      &  -            & 0.58 & 0.99 & 0.9 \\\hline
       a From Ref.~\cite{madelungSemiconductors1991} &&&&& \\
       b From Ref.~\cite{breyDeformationPotentialsValenceband1987} &&&&& \\
       c From Ref.~\cite{nielsenStressesSemiconductorsInitio1985} &&&&&
\end{tabular}
\caption{Strain parameters used in this work}\label{tab:stiffness}
\end{table}

%


\end{document}